\documentclass[reprint,superscriptaddress,amsmath,amssymb,showkeys,aps,prb]{revtex4-2}

\usepackage{graphicx,xcolor}
\usepackage{amsmath,amsfonts,amssymb}
\usepackage{upgreek}
\usepackage[]{siunitx}
\usepackage{hyperref}
\usepackage{pdfpages}
\usepackage{pgffor}

\makeatletter
\AtBeginDocument{\let\LS@rot\@undefined}
\makeatother

\begin{document}

\title{Vibrational properties of LiNbO$_3$ and LiTaO$_3$ under uniaxial stress}

\author{Ekta Singh*}
\email[email]{: ekta.singh1@tu-dresden.de}
\affiliation{Institut f{\"u}r Angewandte Physik, Technische Universit{\"a}t Dresden, 01062 Dresden, Germany}

\author{Mike N. Pionteck*}
\email[email]{: mike.pionteck@theo.physik.uni-giessen.de}
\affiliation{Institut f{\"u}r Theoretische Physik and Center for Materials Research (LaMa), Justus-Liebig-Universit{\"a}t Gießen, 35392 Gie{\ss}en, Germany}

\author{Sven Reitzig}
\affiliation{Institut f{\"u}r Angewandte Physik, Technische Universit{\"a}t Dresden, 01062 Dresden, Germany}

\author{Michael Lange}
\affiliation{Institut f{\"u}r Angewandte Physik, Technische Universit{\"a}t Dresden, 01062 Dresden, Germany}

\author{Michael R{\"u}sing}
\affiliation{Institut f{\"u}r Angewandte Physik, Technische Universit{\"a}t Dresden, 01062 Dresden, Germany}

\author{Lukas M. Eng}
\affiliation{Institut f{\"u}r Angewandte Physik, Technische Universit{\"a}t Dresden, 01062 Dresden, Germany}
\affiliation{ct.qmat: Dresden-W{\"u}rzburg Cluster of Excellence—EXC 2147, TU Dresden, 01062 Dresden, Germany}

\author{Simone Sanna}
\affiliation{Institut f{\"u}r Theoretische Physik and Center for Materials Research (LaMa), Justus-Liebig-Universit{\"a}t Gießen, 35392 Gie{\ss}en, Germany}

* The authors contributed equally to this paper.

\date{\today}

\begin{abstract}
Structural strain severely impacts material properties such as the linear and non-linear optical response. Moreover, strain plays a key role, e.g., in the physics of ferroelectrics and in particular of their domain walls. $\upmu$-Raman spectroscopy is a well-suited technique for the investigation of such strain effects, as it allows to measure the lattice dynamics locally. However, quantifying and reconstructing strain fields from Raman maps requires knowledge on the strain dependence of phonon frequencies. In this work, we have analyzed both theoretically and experimentally the phonon frequencies in the widely used ferroelectrics lithium niobate and lithium tantalate as a function of uniaxial strain via density functional theory and $\upmu$-Raman spectroscopy. Overall, we find a good agreement between our \emph{ab initio} models and the experimental data performed with a stress cell. The majority of phonons show an increase in frequency under compressive strain, while the opposite is observed for tensile strains. Moreover, for E-type phonons, we observe the lifting of degeneracy already at moderate strain fields (i.e. at $\pm\SI{0.2}{\%}$) along the x and y directions. This work hence allows for the systematic analysis of 3D strains in modern-type bulk and thin-film devices assembled from lithium niobate and tantalate.
\end{abstract}


\maketitle
\section{Introduction}

Lithium niobate (LN) and Lithium tantalate (LT) are cornerstone materials in nonlinear optics and optoelectronics, with established applications ranging from second-harmonic generation \cite{Miller1997,Mouras2001}, over optical ring resonators, to holographic memory devices \cite{Yariv1996}, to name just a few. These devices are based on bulk single crystal platforms and have been constantly optimized over the last decades via meticulous analyses of crystal compositional aspects \cite{Iyi1992, Bartasyte2017, Li2015}, degradation effects \cite{Furukawa2000}, defect distribution, and the precise imaging of domain structures \cite{stone12, Nataf2016, Rusing2018a}. Single crystalline LN (or LT) thin films on insulating substrates have recently opened up a completely new scenario. This new platform allows for much higher integrability into existing nanoelectronic and nanophotonic structures \cite{Wang2019,ruesing2019} and outperforms bulk-based devices due to the submicron-scale confinement of optical modes \cite{Zhu2021, Vazimali2022, Wang2018}.

Nevertheless, the transition towards thin-film LN (TFLN) devices emphasizes the importance of crystal properties that so far have played a minor role in LN bulk investigations and device fabrication. Among these, we count the increased influence of mechanical stress ($\sigma$) fields \cite{Takigawa2019,Prabhakar2021}, e.g., due to the bonding process of the LN thin film onto the substrate, or occurring as a result of the poling procedure or waveguide fabrication, which is paramount for the reliable and reproducible production of highly efficient nonlinear optical devices. Accumulated stress does not only increase the chance of critical failure (e.g. due to delamination) \cite{Takigawa2019,Prabhakar2021}, but also influences optical key properties such as refractive indices \cite{Friedrich2017,Zisis2016} and thus endangers the optical responses of a periodically-poled domain grid \cite{Reitzig2021}.

Among established imaging techniques applied for the analysis of LN platforms, Raman spectroscopy can provide critical information about internal mechanical stress. To discriminate the influence of stress fields from other factors, e.g. electrical fields \cite{Stone2011} or compositional heterogeneity \cite{Fontana2015}, and to quantify occurring stress markers, it is essential to gain fundamental knowledge on the relation of phonon properties and mechanical stress fields. To date, the only investigations in this regard were conducted via hydrostatic pressure-cells \cite{MendesFilho1984,jayaraman1986}. However, such experiments only investigate the influence of isotropically applied stress. In reality, effects like stress fields in the vicinity of LN DWs, or mechanical stress in LN thin films, are not isotropic \cite{Jach2004,Scrymgeour2005}. Under these environments, the directionality of stress largely changes the materials properties. Therefore, it is necessary to theoretically predict and experimentally verify the influence of uniaxial stress on phonon properties. This then provides a fundamental reference for future investigations of mechanical stress fields, in analogy to the reference work of Stone \emph{et al.} \cite{Stone2011} for incident electrical fields.

In this work, we firstly provide the theoretical prediction on the evolution of the phonon properties in LN and LT crystals upon uniaxial compression along all three crystal axes via density functional theory (DFT) in Sec.~\ref{sec:methodology_theory}. Secondly, we combine a piezo-driven uniaxial stress cell with a $\mu$-Raman spectroscopy setup to enable the experimental study of stress on the LN crystals. Experimentally, we compare stoichiometric and 5$\%$ MgO-doped LN samples compressed along the x,y, and z axes [see Sec.~\ref{sec:experimental_investigation}]. Sec.~\ref{sec:discussion} then provides a discussion of the calculated and measured data. 

With this work, we thus provide a basis for the quantitative, local stress analysis in LN and LT platforms via $\mu$-Raman spectroscopy and give a reference for the deeper understanding and further optimization of LN- and LT-based devices.

\section{Methodology}

\subsection{Computational Approach}\label{sec:methodology_theory}

The DFT calculations are performed within the generalized gradient approximation (GGA) \cite{Perdew1993} in the formulation of Perdew, Burke, and Ernzerhof (PBE) \cite{Perdew1996} as implemented in the Vienna Ab Initio Simulation Package (VASP) \cite{Kresse1996}. Thereby, projector augmented wave (PAW) \cite{Blochl1994} potentials with projectors up to $l=3$ for Nb and Ta, and $l=2$ for Li and O have been used, which proved to be accurate enough as demonstrated in a previous work \cite{Sanna2015}. The electronic wave functions are expanded into a plane-wave basis set up to kinetic energy of $\SI{475}{eV}$. In order to model uniaxial stress, the lattice constants of the rhombohedral unit cell were adjusted according to the expression

\begin{equation}
    \boldsymbol{a}_n(\mathrm{\epsilon})=\left(\boldsymbol{I}+\mathrm{\epsilon}\right)\boldsymbol{a}_n(0)\mathrm{,}
\end{equation}

where $\mathrm{\epsilon}$, $\boldsymbol{I}$, $\boldsymbol{a}_n(0)$, and $\boldsymbol{a}_n(\mathrm{\epsilon})$ denote the strain tensor, the identity matrix, and the nth unstrained and strained lattice vector, respectively. The strain tensor was then calculated using the elastic compliance tensor $S$ from Ref.~\cite{Warner1967a},

\begin{equation}
    \mathrm{\epsilon}=S\mathrm{\sigma}\mathrm{,}
\end{equation}

\noindent and the boundary conditions for uniaxial stress. This means that all components of the stress tensor $\mathrm{\sigma}$ are set to zero, except for the component in the direction of which uniaxial stress is to be exerted. For instance, if uniaxial stress is to be calculated along the x direction, $\mathrm{\sigma}_1\neq 0=\mathrm{\sigma}_2=\mathrm{\sigma}_3=\mathrm{\sigma}_4$ (Voigt notation) is set as boundary condition. Due to constraints of the unit cell sizes, only stoichiometric LN and LT is calculated. However, as later demonstrated by experiments by comparing 5\% MgO-doped and stoichiometric LN, it is shown that the results for 5\% MgO-doped and stoichiometric LN are comparable, therefore this assumption is justified, to only calculate stoichiometric LN and LT.

The calculations were performed in two ranges for LN and LT: First, the effects of strain, i.e. compressive as well as tensile strain, on the phonon frequencies have been calculated in fine increments up to values reachable in the experiments. Second, calculations for higher values of strains, but more coarse increments, were performed. For LN and LT this procedure has been performed for lower strains up to $\SI{0.10}{\%}$, as well as for higher values up to $\SI{2.4}{\%}$. Here, increments of $\SI{0.02}{\%}$ and $\SI{0.4}{\%}$ have been chosen, respectively. The Hellmann-Feynman forces are minimized under a threshold value of $\SI{0.005}{eV/\text{\AA}}$ by relaxation of the atomic positions. For our calculations of unstrained structures we have used commonly accepted experimental lattice constants ($a_{\text{R}}^{\text{LN}}=\SI{5.494}{\text{\AA}}$, $\alpha^{\text{LN}}=\SI{55.867}{^\circ}$ and $a_{\text{R}}^{\text{LT}}=\SI{5.474}{\text{\AA}}$, $\alpha^{\text{LT}}=\SI{56.171}{^\circ}$) \cite{Sanna2015}.
\begin{table*}
\caption{Description of the samples used for the experimental stress measurement analysis.}
\label{tab:samples}
\begin{ruledtabular}
\begin{tabular}{ c c c c}
\hline
sample name & description & size & stress axis \\
\hline
sLN-x & z-cut stoichiometric LN &  y $\times$ z: 580 $\times$ 113 $\upmu$m$^2$ & x\\
sLN-y & z-cut stoichiometric LN&x $\times$ z:  580 $\times$ 113 $\upmu$m$^2$& y\\
5Mg-LN-x&  z-cut 5$\%$-MgO doped LN & y $\times$ z: 500 $\times$ 140 $\upmu$m$^2$ & x \\
5Mg-LN-y & z-cut 5$\%$-MgO doped LN & x $\times$ z: 400 $\times$ 100 $\upmu$m$^2$ & y\\ 
5Mg-LN-z & x-cut 5$\%$-MgO doped LN & y $\times$ x: 400 $\times$ 100 $\upmu$m$^2$ & z
\end{tabular}
\end{ruledtabular}
\end{table*}

The $\Gamma$-centered phonon frequencies and eigenvectors are derived by the frozen-phonon method \cite{Schmidt1995} without symmetry constraints. For the calculation of the Hessian matrix, atomic displacements of $\SI{0.015}{\text{\AA}}$ in each Cartesian direction are considered. Since our approach does not take into account the long-range electric fields accompanying the longitudinal-optical (LO) phonons, our calculations are restricted to transversal-optical (TO) phonons, only. A $8\times8\times8$ k-point mesh is used to sample the first Brillouin-zone corresponding to the orthorhombic unit cell, which yields 192 irreducible points.

\subsection{Experimental Procedure}\label{sec:experimental_investigation}
In order to complement the above theoretical investigations, we measured the frequency shift of the phonon modes in LN by combining $\upmu$-Raman spectroscopy with a uniaxial stress cell \cite{Singh2022}. All measurements were performed using a LabRAM HR spectroscope from Horiba in 180$^\circ$ back-scattering geometry. In this setup, a HeNe laser at $\SI{632.8}{\mathrm{nm}}$  wavelength was focused onto the samples using a 100X and 0.9 NA objective. For each data point, two acquisitions were collected for $\SI{10}{s}$ each, while using a diffraction grating of 1800 grid lines per millimeter for spectral analysis. For these settings, the corresponding spectral resolution is $\SI{0.56}{ cm^{-1}}$, which was determined as part of the calibration using a He-Ne gas discharge lamp. Here, the plasma line at a wavelength of 671.7 nm has a natural line width of 0.025 nm ($\SI{0.56}{ cm^{-1}}$), which corresponds to the spectral resolution i.e. the standard deviation, which only means how well the two overlapping peaks can be resolved. On the other hand, with the help of the standard error obtained from peak fitting, each peak center can be pinpointed with a 100-1000 times more precision, and this is the quantity we are interested in.

All the samples used in this work were firmly attached to the uniaxial stress cell using black stycast epoxy resin \cite{Stycast2014} such that the strain is fully transmitted to the sample through the epoxy resin. The force applied to the sample was directly measured through a force sensor attached to the back of the device. This force was later converted to stress using the relation
\begin{equation}
 \sigma = \dfrac{F}{A},  
\end{equation}
where $A$ denotes the cross-section of the samples provided in table~\ref{tab:samples}, and $F$ is the force measured by the force sensor. In order to compensate for any charge-induced effects due to the piezoelectric effect, the top and bottom faces of the sample surfaces were electrically grounded using $\SI{10}{\mathrm{nm}}$-thick chromium electrodes. All the measurements were always performed across the chromium electrodes, which at the given thickness are still sufficiently transparent and do not affect our measurements, as also shown in the past for \textit{in situ} second-harmonic microscopy studies \cite{kirbus2019}.

A total of five samples were measured [see Tab.~\ref{tab:samples}]. First, stoichiometric LN, being close to the ideal crystals considered for the theoretical calculations, is measured and used to compare theory with experiment. Second, 5$\%$ MgO-doped LN samples are chosen, because they are widely used among the ferroelectrics and integrated optics community due to their low photorefraction and highly conductive domain walls \cite{ruesing2019,Godau2017,Bednyakov2018,Werner2017}. Comparing the results on the stoichiometric and MgO-doped crystals allows us to see whether or not there are any stoichiometry-specific changes in the responses, which would also not be covered by the DFT calculations.
 
 \subsubsection{Data extraction method}
The Raman spectra of single-crystalline LN and LT are well understood and have been investigated thoroughly before \cite{Sanna2015, Bartasyte2019,Rusing2016,Margueron2012, Irzhak2022}. The optical phonons of LN and LT consist of four A$_1$, five A$_2$, and nine doubly-degenerated E phonon branches. Here, A$_1$ and E are optical branches and are both Raman and Infrared active. In contrast, the A$_2$ branch is optically inactive. Furthermore, depending on the scattering geometry these branches either appear as TO or LO modes. Conveniently LN and LT peaks are labeled by type of symmetry (A$_1$ or E), followed by the specific type (TO or LO) and consecutive number counting from low to high frequency. For example, the phonon E(TO$_9$) labels the ninth, i.e. highest frequency, E-type TO phonon. The assignment of peaks in this work has been performed according to the most recent accepted assignment by Margueron \emph{et al.}, Sanna \emph{et al.} and R\"{u}sing \emph{et al.} \cite{Rusing2016,Margueron2012,Sanna2015}.

\begin{figure}
	\includegraphics{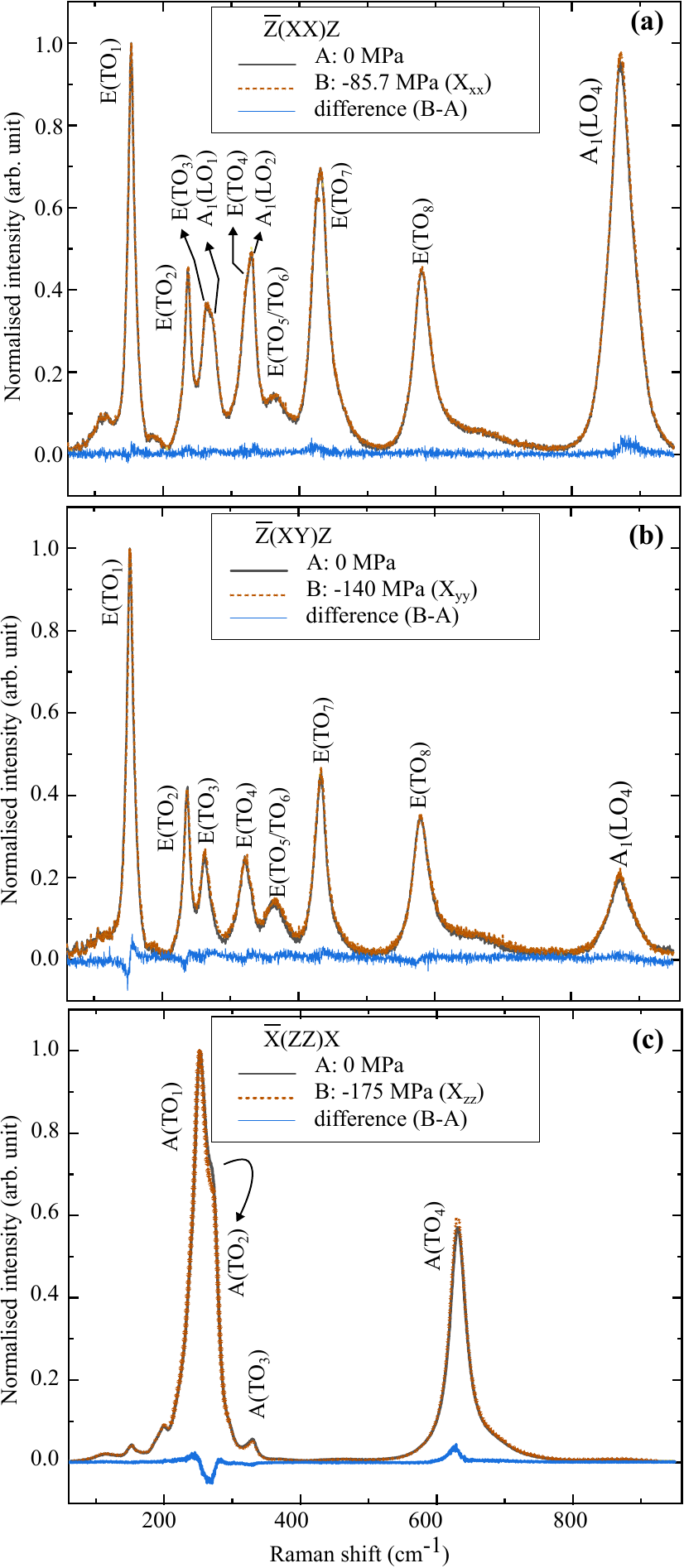}
	\caption{\label{fig:data_extraction}(a) $\mathrm{\overline{Z}}$(XX)Z Raman spectra of unstressed (0 MPa) and x compressed (-85.7 MPa) 5$\%$ MgO-doped LN (5Mg-LN-x).  (b) $\mathrm{\overline{Z}}$(XY)Z Raman spectra of unstressed (0 MPa) and y compressed (-140.7 MPa) 5$\%$ MgO-doped LN (5Mg-LN-y). (c)  $\mathrm{\overline{X}}$(ZZ)X Raman spectra of unstressed (0 MPa) and z compressed (-175 MPa) z-axis of 5$\%$ MgO-doped LN (5Mg-LN-z). The difference curve in all (a)-(c) graphs represents the difference spectra between the stressed and unstressed states of the sample.}
\end{figure}

\begin{figure*}
	\includegraphics{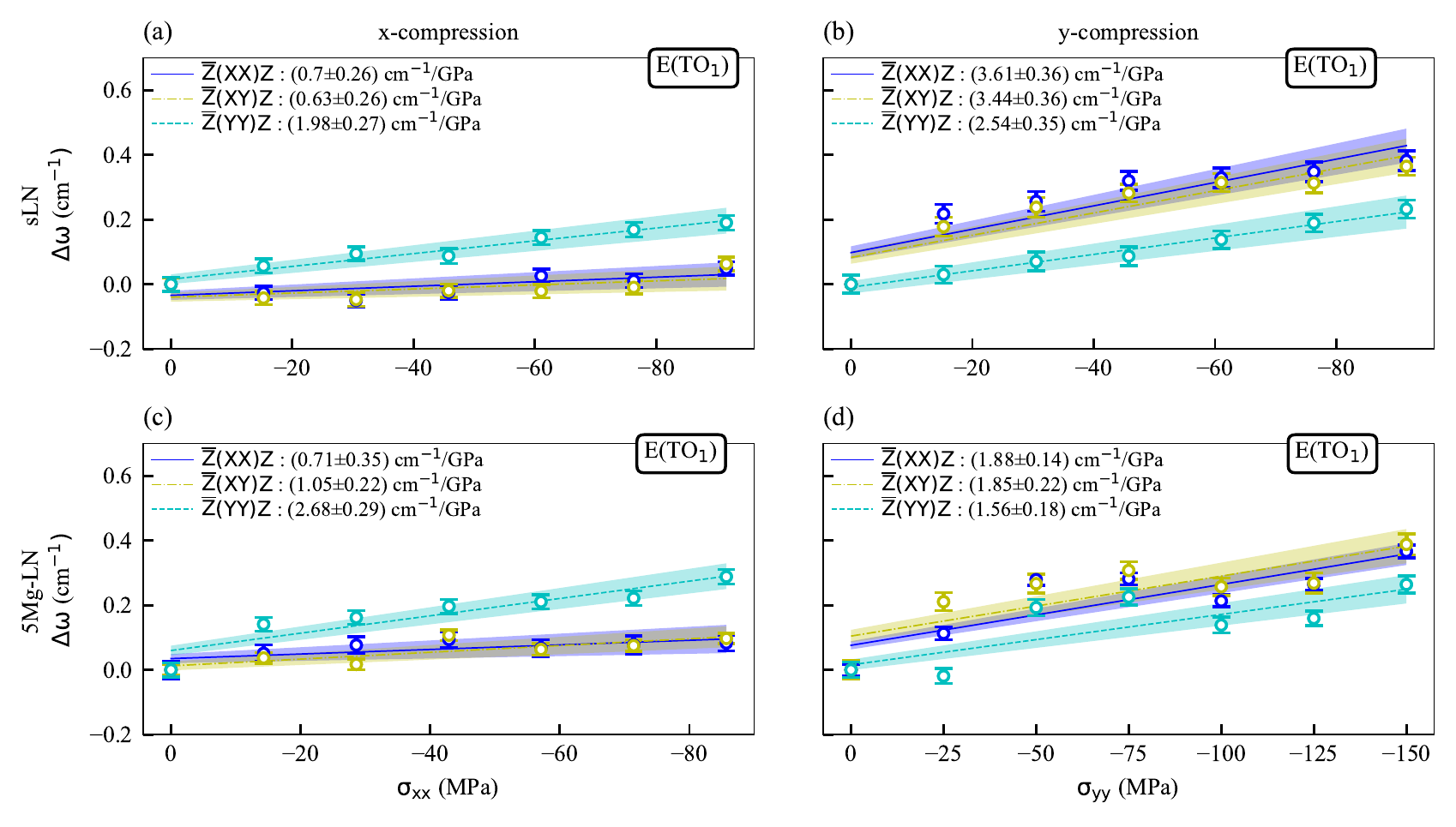}
	\caption{\label{fig:result_1} Measured response of the E(TO$_1$) phonon mode (a)-(b) in the samples sLN compressed along the x and y axes, respectively and (c)-(d) in the sample 5Mg-LN compressed along the x and the y axes, respectively. The stoichiometric and 5$\%$ MgO-doped LN behave similarly.}
\end{figure*}

Figure~\ref{fig:data_extraction}(a), (b) and (c) shows example Raman spectra of the 5Mg-LN-x, 5Mg-LN-y, and 5Mg-LN-z samples, respectively. These spectra are measured in different scattering geometries represented by the Porto notation $k_i(e_i,e_s)k_s$, where $k_i$ and $k_s$ represent the direction of incident and scattered light in crystal coordinates, and $e_i$ and $e_s$ denote the direction of polarization of both incoming and Raman scattered beams, respectively \cite{Rusing2016}. Here, each plot contains 3 spectra, where spectrum A belongs to the unstressed ($\SI{0}{\mathrm{MPa}}$) state of the sample while spectrum B describes a compressed state of the sample. The third is the difference spectrum of the former two. To eliminate the intensity influences in the difference spectrum and to solely highlight the effect of the frequency shift due to applied stress, all the spectra were first normalized individually by subtracting the noise background (if applicable) and divided by the intensity of the highest peak \cite{Rusing2018a}. From the different spectra of Fig.~\ref{fig:data_extraction}(b), it can be clearly noticed that the frequency of the E(TO$_1$) phonon mode in $\mathrm{\overline{Z}}$(XY)Z geometry increased under compression, while the frequency for A$_1$(TO$_1$) in $\mathrm{\overline{X}}$(ZZ)X geometry decreased. Apart from these large differences, further changes are difficult to notice. So, to quantify frequency shifts as a function of stress, the fitting of the spectra by Lorentzian functions is performed. In this way, the frequencies at the maxima of the phonon peaks were determined for different stress values and plotted against applied stress, such as shown in Fig.~\ref{fig:result_1}. Via statistical analysis of the phonon shifts, we are able to deduce the experimental phonon-stress-shift coefficients with a precision of 0.01 cm$^{-1}$/$\%$. 

Additionally, it is well-known \cite{Sidorov2016,Ridah1997} that the spectra of congruent or MgO-doped LN show broader and less resolvable peaks as compared to the stoichiometric LN. Here, various peaks overlap with each other, that then leads to a more challenging peak assignment. For example, the pairs E(TO$_3$)- A$_1$(LO$_1$), A$_1$(LO$_2$)-E(TO$_4$), and A$_1$(LO$_3$)-E(TO$_7$) overlap with each other in the $\mathrm{\overline{Z}}$(XX)Z and $\mathrm{\overline{Z}}$(YY)Z geometry of MgO-doped LN. Therefore, in Sec.~\ref{sec:experimental_results}, to avoid any misleading conclusions, for these peaks the Raman shifts with the stress have not been fitted [see Table~\ref{table:mgo_sln} and \ref{table:compare_new}]. 

Furthermore, to find an optimum location on the sample, a series of optimization measurements [see SI for further information] are performed. As a result, all the measurements were performed in the center of the sample and 10 $\upmu$m below the sample surface. 

\section{Results}

\subsection{Experimental Results}\label{sec:experimental_results}
For the sake of simplicity, we only show the pressure dependency for selected phonon modes; the complete results are shown in the Supplement information (SI).

Fig.~\ref{fig:result_1} shows the response of the E(TO$_1$) phonon mode under x and y compression for both the stoichiometric and 5$\%$-MgO doped LN samples. The three curves in a single plot represent the three different scattering geometries given in Porto's notation. Since the change of the frequency ($\Delta \omega$) with respect to the unstressed state is the quantity of interest, we have plotted the frequencies with respect to the $\SI{0}{\mathrm{MPa}}$ value as a function of stress. Hence, we also account for any offset due to sample mounting or bending effects as discussed in the SI. 

The coloured region around the fitted lines show the one-sigma confidence interval, calculated from linear fitting which included the standard error of individual peaks. The E(TO$_1$) mode shows a linear positive slope for the compression in all cases. In the case of x compression for both sLN-x and 5Mg-LN-x samples, the slope is larger in $\mathrm{\overline{Z}(YY)Z}$ measurement geometry. On the other hand, when the same samples are compressed along the y crystallographic axis, the slope is larger for $\mathrm{\overline{Z}(XY)Z}$ and $\mathrm{\overline{Z}(XX)Z}$ scattering geometries. We have observed this behavior for almost all the phonon modes [see SI]. The value of different slopes in different measurement geometry can be interpreted as the lifting of the two-fold degeneracy of the E-modes predicted by theory as discussed below. 

When comparing 5Mg-LN-x/-y and sLN-x/-y, the slopes are similar as shown in Fig.~\ref{fig:result_1}. The origin of the small deviations could be due to the presence of defects in doped-LN. This argument is consistent with the work done by Tejerina \emph{et al.}, where they show that doping increases the phonon shift under hydrostatic pressure ($\Delta \omega$/$ \Delta \sigma$)~\cite{TEJERINA2013581}, especially for the A$_1$(TO$_2$), A$_1$(TO$_3$), and E(TO$_4$) phonon modes. The comparison for other peaks is also summarized in Table~\ref{table:mgo_sln}. 

As will later be shown by theory, the phonon frequency should decrease under tension with a slightly different magnitude. That is why the phonon response of sample sLN-x is also measured experimentally in $\mathrm{\overline{Z}(YY)Z}$ scattering geometry under tension. For tensile strain, the opposite response is observed as expected: the frequency decreased on applying tension to the sample. For example, phonon mode E(TO$_6$) shows the largest response as for compression with a slope of $\mathrm{-8.15}$ $\pm$ {2.88} cm$^{-1}$/GPa. The plot for E(TO$_6$) and the slopes for all peaks are provided in SI.

\begin{table*}
	\centering
	\caption{\label{table:mgo_sln} Experimentally determined Raman shift changes $\Delta \omega$ as a function of stress $\sigma$ for x compression. The slopes were obtained after linearly fitting the experimental changes in frequency upon x compression. An asterisk labels silent modes \cite{Rusing2016}. In the case of 5$\%$ MgO-doped LN, many peaks overlap with each other [see Fig.~\ref{fig:data_extraction}(a)-(b)], therefore extracting the peak frequency with an acceptable error was not possible. This is represented by the '-' sign.}
	\begin{ruledtabular}
	
	\begin{tabular}{c|c c c|c c c}
	\hline
    &\multicolumn{3}{c|}{sLN-x (cm$^{-1}$/GPa)}&\multicolumn{3}{c}{5Mg-LN-x (cm$^{-1}$/GPa)}\\[1ex]

    Phonon modes&$\mathrm{\overline{Z}}$(XX)Z & $\mathrm{\overline{Z}}$(XY)Z & $\mathrm{\overline{Z}}$(YY)Z &$\mathrm{\overline{Z}}$(XX)Z & $\mathrm{\overline{Z}}$(XY)Z & $\mathrm{\overline{Z}}$(YY)Z\\
	\hline
	$\mathrm{A_1(LO_1)}$ & 2.13 $\pm$ 0.52 & * & 2.06 $\pm$ 0.58 & - & * & -\\
    $\mathrm{A_1(LO_2)}$ & 3.29 $\pm$ 0.17 & * & 3.66 $\pm$ 0.17 & - & * & -\\
    $\mathrm{A_1(LO_3)}$ & 0.39 $\pm$ 0.94 & * & 0.29 $\pm$ 0.71 & - & * & -\\
    $\mathrm{A_1(LO_4)}$ & 1.96 $\pm$ 0.21 & * & 2.32 $\pm$ 0.19 & 3.07 $\pm$ 0.83 & * & 3.79 $\pm$ 0.80\\[1ex]
    \hline
    $\mathrm{E(TO_1)}$ & 0.70 $\pm$ 0.26 & 0.63 $\pm$ 0.26 & 1.98 $\pm$ 0.27 & 0.71 $\pm$ 0.35 & 1.05 $\pm$ 0.22 & 2.68 $\pm$ 0.29\\
    $\mathrm{E(TO_2)}$ & 0.34 $\pm$ 0.17 & 0.26 $\pm$ 0.25 & 1.68 $\pm$ 0.16 & -0.51 $\pm$ 0.45 & -0.63 $\pm$ 0.52 & 1.52 $\pm$ 0.3\\
    $\mathrm{E(TO_3)}$ & 1.28 $\pm$ 0.44 & 1.39 $\pm$ 0.65 & 1.67 $\pm$ 0.42 & - & - & -\\
    $\mathrm{E(TO_4)}$ & 2.12 $\pm$ 0.35 & 1.56 $\pm$ 0.59 & 3.02 $\pm$ 0.4& - & - & -\\
    $\mathrm{E(TO_{5/6})}$ & 6.36 $\pm$ 0.98 & 6.25 $\pm$ 1.5 & 8.17 $\pm$ 0.87& 5.87 $\pm$ 4.23 & 3.77 $\pm$ 3.69 & 9.38 $\pm$ 2.16\\
    $\mathrm{E(TO_7)}$ & 0.44 $\pm$ 0.37 & 0.86 $\pm$ 1.51 & 1.24 $\pm$ 0.22 & - & - & -\\
    $\mathrm{E(TO_8)}$ & -0.47 $\pm$ 0.30 & 0.37 $\pm$ 0.36 & 3.03 $\pm$ 0.27 & -0.29 $\pm$ 0.8 & 1.66 $\pm$ 0.69 & 4.32 $\pm$ 0.67\\
    \hline
	\end{tabular}

	\end{ruledtabular}
\end{table*}

\subsection{Strain Simulation}\label{sec:theoreticalResults}

First, to assess the quality of the models in this work, the calculated phonon frequencies of unstrained stoichiometric LN and LT are compared with previous works \cite{Sanna2015} (all data appear in the SI). Compared to our experiment, the largest deviation from the experimental values for LN and LT occurs for the A$_1$(TO$_4$) mode with $\SI{20}{\mathrm{cm}^{-1}}$ and $\SI{22}{\mathrm{cm}^{-1}}$, respectively. The mean deviation from experimental values for LN and LT is about $\SI{8.7}{\mathrm{cm}^{-1}}$ and $\SI{8.4}{\mathrm{cm}^{-1}}$. Thus, compared with the mean deviation of the theoretical results of Ref. \cite{Sanna2015} which was about $\SI{10.7}{cm^{-1}}$ and $\SI{10.8}{cm^{-1}}$ for LN and LT, there is a close agreement with both the theoretical and experimental values. In addition, the typical deviations are also consistent with literature values, such as Refs. \cite{Rusing2016,Hermet2007}.

\begin{figure}
		\centering
		\includegraphics{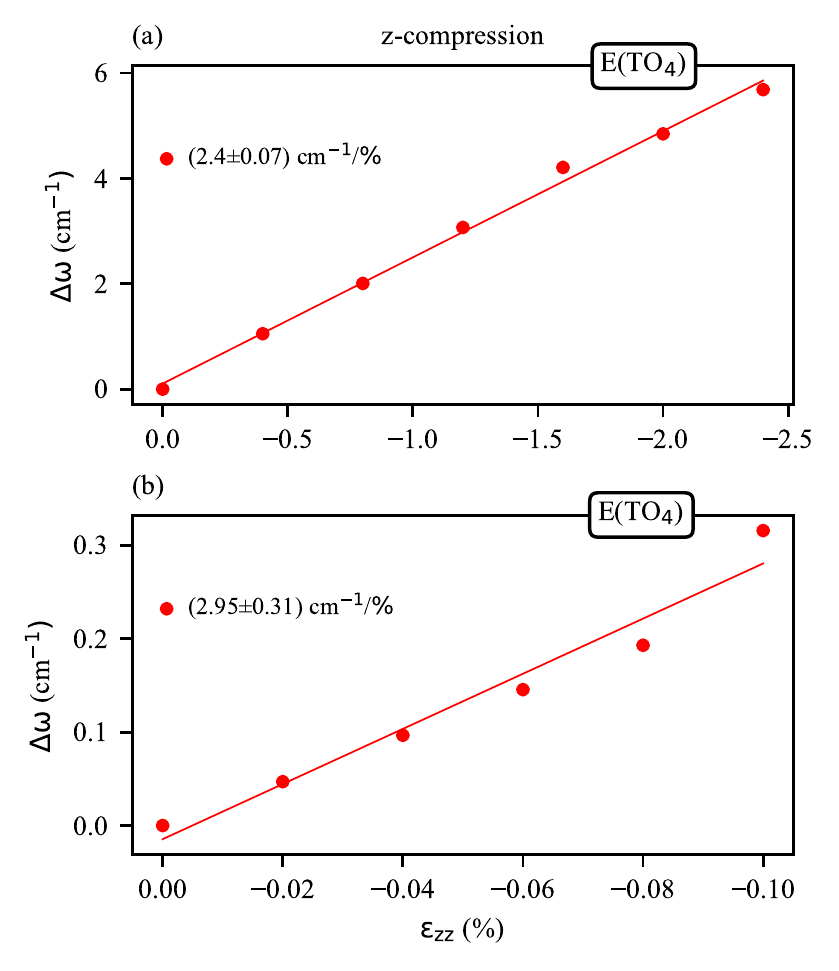}
		\caption{DFT calculated relative frequency shifts of the E(TO$_4$) mode of stoichiometric LN as a function of strain in z-direction in (a) steps of $\SI{0.4}{\%}$ and (b) steps of $\SI{0.02}{\%}$. The linear fit and the standard deviation of both are plotted and their parameters displayed. The slopes for strain are defined as $\mathrm{(\Delta\omega)}/(\mathrm{|\Delta\epsilon|)}$ and displayed with the fitting error.}
		\label{fig:error}
\end{figure}

In order to demonstrate the dependence of the phonon frequencies predicted by our calculations, the E(TO$_4$) mode of LN was chosen [see Fig.~\ref{fig:error}]. Figures ~\ref{fig:error}(a) and (b) show the results of the calculations with lower and higher strains, respectively. The frequency shift of the E(TO$_4$) mode has an approximately linear dependence on the applied compression. Such a linear dependence is observed in all of our calculations for each mode for both the compressive and tensile strain in stoichiometric LN and LT. This is consistent with the results of LN and LT under hydrostatic pressure by Mendes-Filho \emph{et al.} \cite{MendesFilho1984}. In Fig.~\ref{fig:error}(a) the frequency increases up to $\SI{6}{\mathrm{cm}^{-1}}$ and in (b) up to $\SI{0.3}{\mathrm{cm}^{-1}}$.

\begin{figure*}
		\centering
		\includegraphics{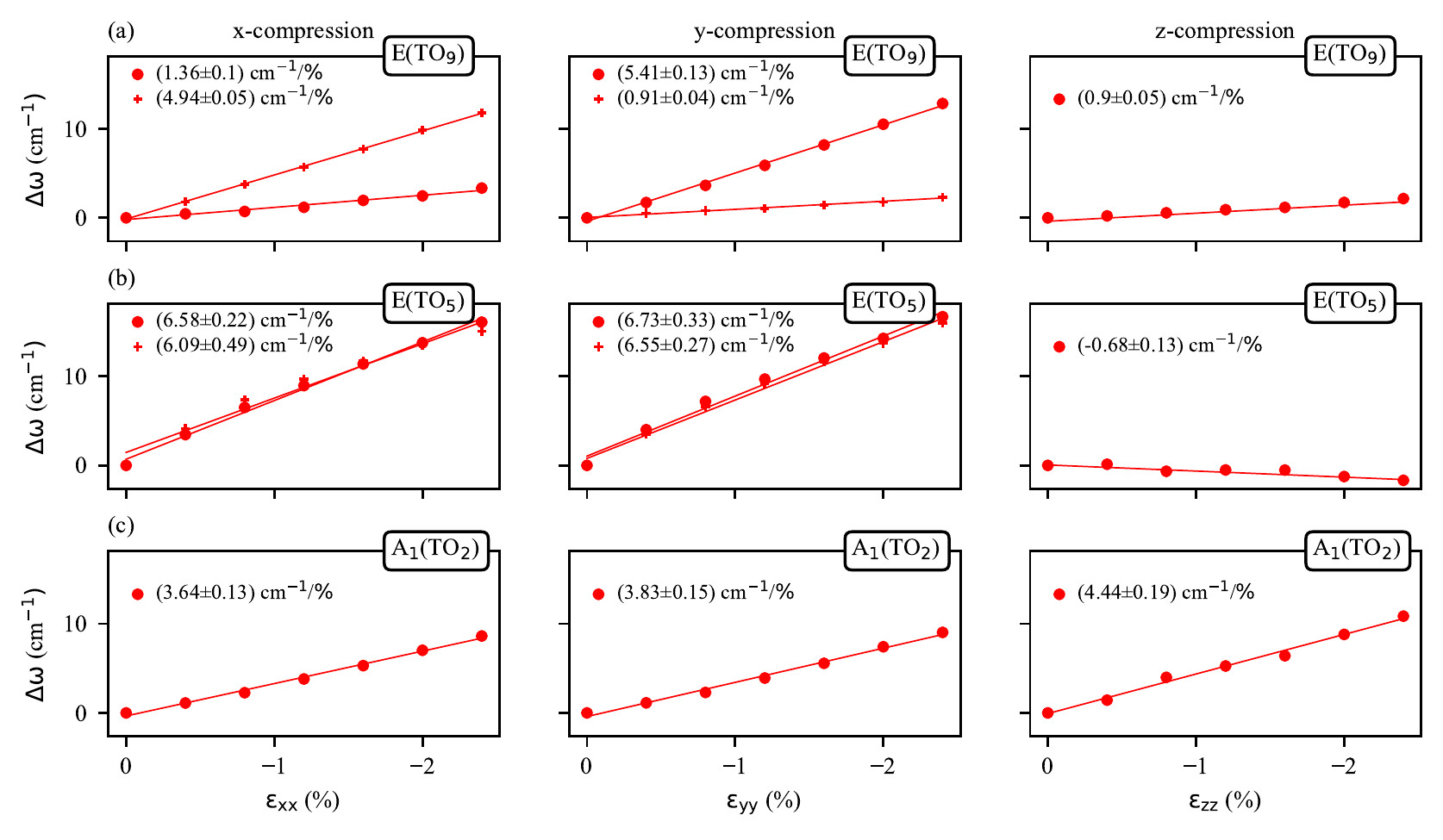}
		\caption{The DFT calculated relative frequency shifts of (a) E(TO$_9$) (first row), (b) E(TO$_5$) (second row) and (c) A$_1$(TO$_2$) (third row) modes of stoichiometric LN as a function of strain in steps of $\SI{0.4}{\%}$ along the x, y and z direction. The slopes for strain are defined as $\mathrm{(\Delta\omega)}/(\mathrm{|\Delta\epsilon|)}$ and displayed with the fitting error.}
		\label{fig:theory}
	\end{figure*}

Our calculations predict for stoichiometric LN and LT that the degeneracy of the E modes is lifted under strain applied along the x and y direction due to the reduction of the threefold symmetry. This results in splitting into two branches, which is consistent with previous studies on various 2D materials, that reported a mode splitting under strain due to the reduction of symmetry \cite{Doratotaj2016,Huang2009}. Consequently, for strains applied along the x and y direction, all E modes are considered separately. For example, the E(TO$_9$) mode for compression in x and y directions has a significant splitting [see Fig.~\ref{fig:theory}(a)]. In contrast, it is noticeable that the E(TO$_9$) mode remains degenerate for z compression since the considered strains in the z direction of both LN and LT do not lead to any reduced symmetry. This is the case for all E modes of LN and LT under strain.

Since the slopes, for strains in the x and y directions of LN and LT, are very similar [see SI] due to the symmetry of the crystals, Table~\ref{tab:tabelTheory} shows only the slopes for strain in the x direction. With the exception of the A$_1$(TO$_1$) and A$_1$(TO$_4$) modes, compressive strain leads to an increase in frequency, while tensile strain leads to a decrease in frequencies, which is consistent with results from previous research on other materials \cite{Pak2017,Oliveira2015}. Comparing the slopes under the compressive and tensile strain of LN, we find that they hardly differ for the individual modes. This behavior is not surprising, however, it is not universal, as shown for a 2D material by Pak \emph{et al.} \cite{Pak2017}. Although under both compressive and tensile strain the E(TO$_{5/6}$) modes have the largest frequency shifts, the E(TO$_6$) mode has a larger slope than the E(TO$_5$) mode under compression. However, this relation turns around under tensile strain as can be seen in Table~\ref{tab:tabelTheory}. This applies to strain in the x and y direction for both LN and LT (see SI).

	 \begin{table}
         \caption{Calculated slopes of transversal A$_1$ and E modes of stoichiometric LN under compressive and tensile strain, and stoichiometric LT under compressive strain in the x-direction at higher strains in $\mathrm{cm}^{-1}$/\%. The slopes for strain are defined as $\mathrm{(\Delta\omega)}/(\mathrm{|\Delta\epsilon|)}$.}
        \label{tab:tabelTheory}
 \begin{ruledtabular}
 \begin{tabular}{ c c c c }
\hline
  Phonon & LN comp. & LN tens. & LT comp.\\
 mode & $\mathrm{\Delta\omega}\mathrm{[cm^{-1}/\%]}$ & $\mathrm{\Delta\omega}\mathrm{[cm^{-1}/\%]}$ & $\mathrm{\Delta\omega}\mathrm{[cm^{-1}/\%]}$\\
\hline

$\mathrm{A_1(TO_1)}$ & 1.31 & -2.87 & 3.16 \\
$\mathrm{A_1(TO_2)}$ & 3.64 & -4.78 & 4.51 \\
$\mathrm{A_1(TO_3)}$ & 5.18 & -7.08 & 4.84 \\
$\mathrm{A_1(TO_4)}$ & 3.95 & -3.79 & 3.93 \\[1ex]
$\mathrm{E(TO_1)}$ & 0.68/3.26 & -1.56/-3.37 & 1.32/1.45 \\
$\mathrm{E(TO_2)}$ & 1.52/0.32 & -3.69/0.04 & 2.68/0.91 \\
$\mathrm{E(TO_3)}$ & 1.09/2.24 & -4.40/-1.56 & 2.62/2.45 \\
$\mathrm{E(TO_4)}$ & 2.19/1.98 & -3.42/-1.65 & 2.73/2.40 \\
$\mathrm{E(TO_5)}$ & 6.58/6.09 & -9.57/-13.57 & 7.39/8.31 \\
$\mathrm{E(TO_6)}$ & 12.84/12.39 & -4.93/-3.76 & 14.82/13.58 \\
$\mathrm{E(TO_7)}$ & 4.18/2.72 & 1.01/-1.52 & 0.83/2.02 \\
$\mathrm{E(TO_8)}$ & 3.63/0.91 & -0.67/-4.51 & 4.65/1.54 \\
$\mathrm{E(TO_9)}$ & 1.36/4.94 & 0.70/-5.12 & 4.00/3.45 \\
\hline

 \end{tabular}
 \end{ruledtabular}
 \end{table}

\begin{figure*}
		\centering
		\includegraphics{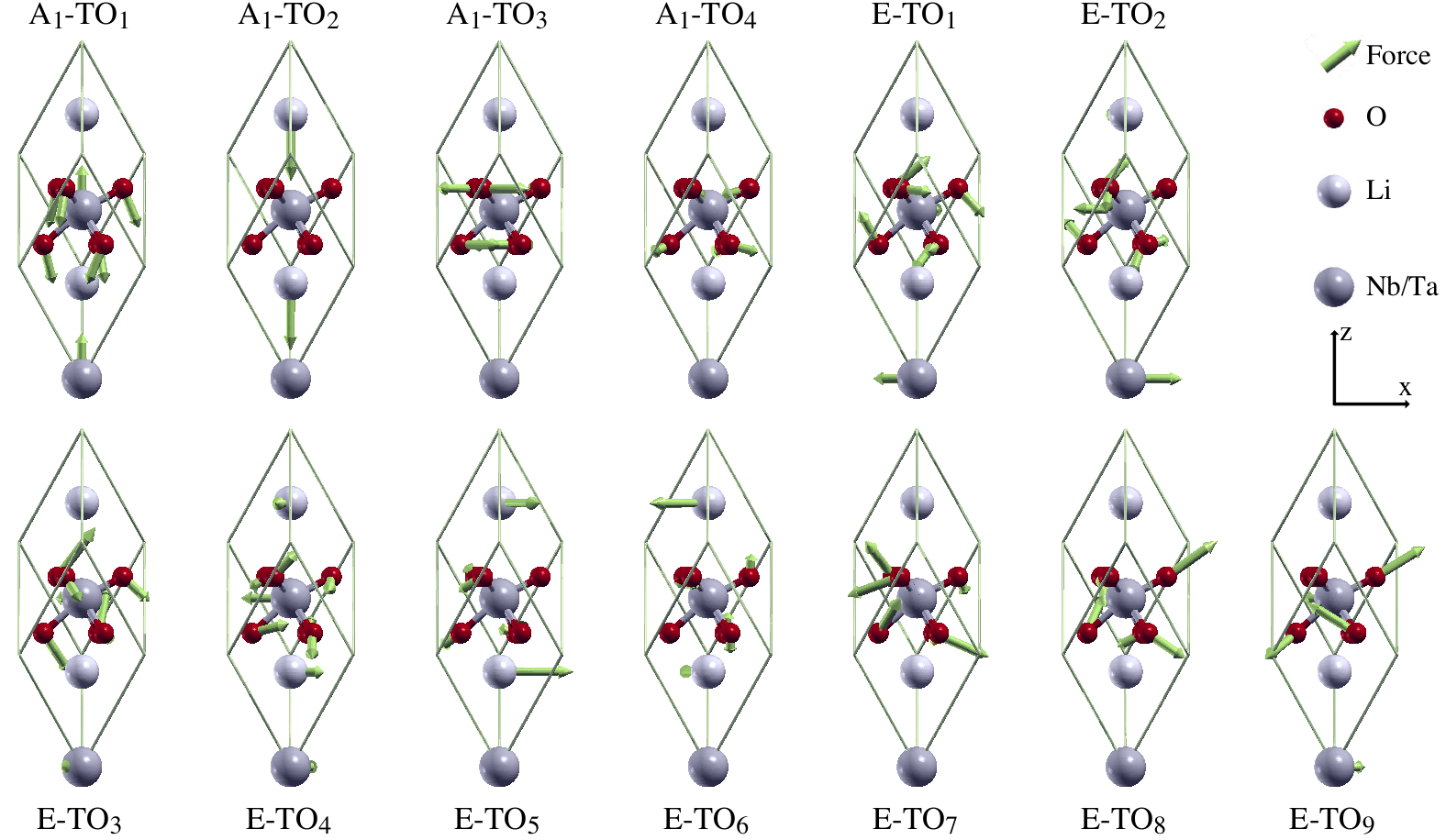}
		\caption{The eigenvectors of the Raman active vibrations of unstrained stoichiometric LN and LT in the rhombohedral unit cell of the transversal phonon modes with A$_1$ and E symmetry. Thereby light grey is denoted as Li, dark grey as Nb/Ta, and red as O ion, while the green arrows represent the displacement direction.}
		\label{fig:ev}
	\end{figure*}

In addition, the results of LT under compression are listed in Table~\ref{tab:tabelTheory} for the same modes. In direct comparison with the results of LN, they are very similar along both directions, as well as magnitude. This is in agreement with Mendes-Filho \emph{et al.} \cite{MendesFilho1984} and expected since LN and  LT are isostructural. Nevertheless, there are some small deviations. In particular, the degeneracy lifting is attenuated compared to LN. For instance, the upper branch of the E(TO$_1$) mode has a notable lower slope than that of LN. Also, the E(TO$_9$) mode exhibits a significantly attenuated degeneracy lift for LT under compression than for LN. These deviations can be explained with the help of the eigenvectors. In Fig.~\ref{fig:ev} the eigenvectors of A$_1$(TO$_{1-4}$) and E(TO$_{1-9}$) are shown in the rhombohedral unit cell \cite{Rusing2016,Sanna2015}. Such modes with high vibrational contributions from Nb and Ta ions, like the E(TO$_1$) mode, lead to a slightly different vibrational behavior \cite{Sanna2015}. The modes that have in turn low contributions of Nb and Ta ions with respect to their eigenvectors, are very similar under compression of LN and LT. Since in the E(TO$_9$) mode, the displacement of O is more involved than that of Nb/Ta, the reason for the difference between LN and LT may be due to the nature of Nb/Ta-O bond, similar to observations in Refs. \cite{Sanna2015,Rusing2016}. The mean Nb-O distance ($\SI{2.02}{\text{\AA}}$) is larger than that of Ta-O ($\SI{2.00}{\text{\AA}}$). Stronger bonding with the oxygen atoms, and thus larger deformation of the oxygen octahedron, might be the reason for the difference between LN and LT concerning the E(TO$_9$) mode. Also, slight differences in E(TO$_7$) and E(TO$_8$), which are pure distortions of the oxygen octahedron, might be related to the different Nb/Ta-O bonds.

The E(TO$_{5/6}$) and A$_1$(TO$_2$) modes are characteristic for LN and LT under strain, since these have particularly high slopes, as shown in Table~\ref{tab:tabelTheory} and Fig.~\ref{fig:theory}(b) and (c). This is also in agreement with Ref. \cite{MendesFilho1984}. The E(TO$_{5/6}$) are especially sensitive to strain in the x and y directions. This can be explained by a simple analysis of their eigenvectors. In Fig.~\ref{fig:ev} it can be seen that these two modes, in contrast to all other modes, are characterized by particularly large displacements of the lithium ions parallel to the x-y-plane. Hence, these modes are most affected when the distances between the ions parallel to the x-y-plane are shortened. The A$_1$(TO$_2$) mode has among the largest slope for strain along the z direction. This can be reasonably explained in terms of the eigenvectors as well. As seen in Fig.~\ref{fig:ev}, the ions of the A$_1$(TO$_2$) mode vibrate in z direction. When compression in the z direction shortens the bond length between the ions parallel to their displacement, the frequency is expected to increase. As the eigenvectors of the A$_1$(TO$_2$) mode point most in the z direction, it consequently has the largest slope compared to all other modes when compressed in the z-direction. All further data and results on LN and LT under compressive and tensile strain are provided in the SI.
\begin{figure*}
	\includegraphics{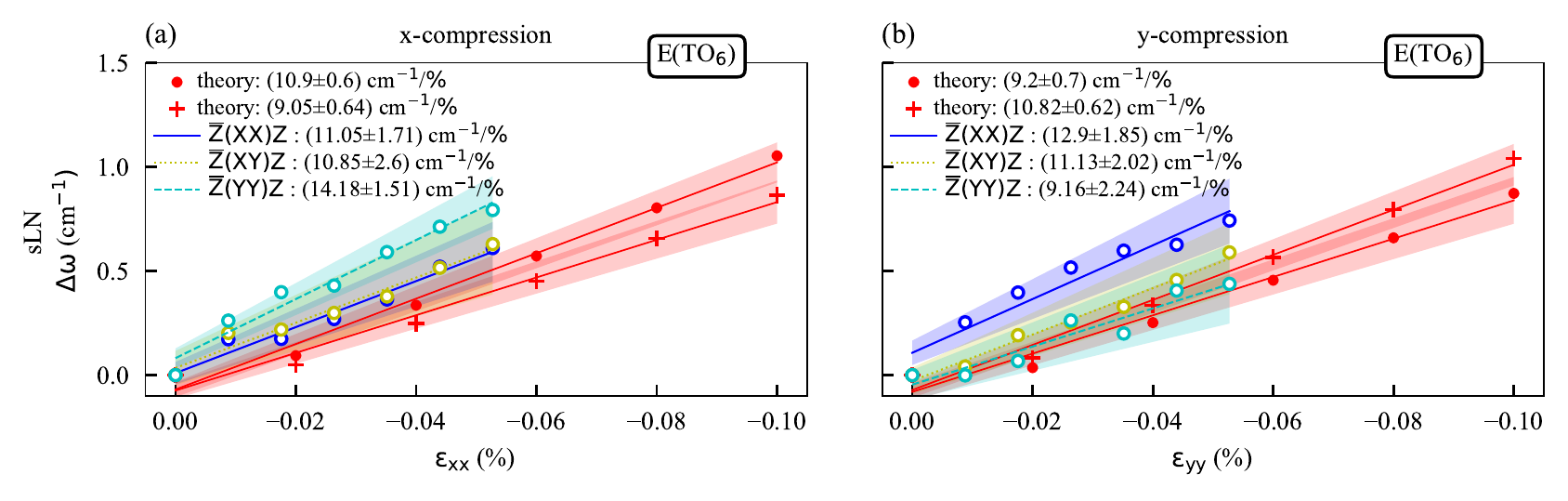}
	\caption{\label{fig:result_compare}Comparison of the experimental and theoretical frequency shift (at lower strains): (a)-(b) E(TO$_6$) phonon mode under compression along the x and y axes of stoichiometric LN. The shaded region around all the curves is the confidence interval of the linear fit.} 
\end{figure*}

\subsection{Discussion: Comparing theory and experiment}\label{sec:discussion}

As mentioned in the introduction section, the calculations are performed in units of strain and the experiments are performed in units of stress. In order to compare both results we have converted the experimental values of stress into values of strain using the equation given below:
\begin{equation}
    \sigma = E \epsilon\mathrm{,}
\end{equation}

\noindent where $E$ is Young's Modulus of a sample \cite{Newnham2004}. The calculated value of Young's Moduli of LN along the x, y, and z axes are $\SI{173.6}{\mathrm{GPa}}$, $\SI{173.07}{\mathrm{GPa}}$, and $\SI{201.0}{\mathrm{GPa}}$, respectively [see calculations in SI]. For every measured stress, we calculated the respective strain value and plotted the corresponding frequency shift along with the theoretical data as a function of strain [see Fig.~\ref{fig:result_compare}].
Since the stoichiometric samples are the closest system to the stoichiometric system considered for theoretical calculations, they are compared with theoretical data in Fig.~\ref{fig:result_compare}.

For x compression, theoretically, the E(TO$_6$) mode shows lifting of the degeneracy in accordance with the results of Sec.~\ref{sec:theoreticalResults}. For a more accurate comparison, the slopes for higher strains for y and z compression are listed in Table~\ref{table:compare_new}. The theoretical results on the degeneracy lifting offer a new interpretation of the experimental results. In particular, comparing the data for the E(TO$_7$) mode with compression in the x direction, it can be seen that the data for the $\mathrm{\overline{Z}}$(YY)Z resembles the upper branch, and its slope of the theoretical data, and the data for the $\mathrm{\overline{Z}}$(XY)Z  and $\mathrm{\overline{Z}}$(XX)Z geometry is similar to the theoretical data and slope of the lower branch. For y compression, the same can be observed, with the $\mathrm{\overline{Z}}$(XY)Z  and $\mathrm{\overline{Z}}$(XX)Z geometries having the higher slope, respectively, and the $\mathrm{\overline{Z}}$(YY)Z having the lower slope, i.e. the slopes flip for y compression. Consequently, it can be assumed that with the help of different scattering geometries the lifting of the degeneracy by x and y compression can be reproduced experimentally.

Based on this interpretation, it can be concluded that the experimental and theoretical data fit very well. For a more accurate comparison, the slopes for higher strains for y and z compression are listed in Table~\ref{table:compare_new}.

Interpreting the experimental data as lifting the degenerate E modes on x and y compression can also help characterize compressed LN. From the data, it can be implied that one has x compressed LN when the $\mathrm{\overline{Z}}$(YY)Z geometry measures a higher slope than for the $\mathrm{\overline{Z}}$(XY)Z  and $\mathrm{\overline{Z}}$(XX)Z geometries. If, on the other hand, one has y compressed LN, this arguement just reverts. In contrast, if no splitting E modes are observed in different scattering geometries, it can be concluded that the LN under investigation must be a z-compressed sample.  Furthermore, the experimental comparison in Fig.~\ref{fig:result_compare} and Table~\ref{table:compare_new}, confirms the prediction that the E(TO$_6$) mode is the mode with highest slope for x and y compression. In comparison with Sec.~\ref{sec:experimental_results} and \ref{sec:theoreticalResults}, the slopes for x compression are very similar to those for y compression.

\begin{table*}
	\centering
	\caption{\label{table:compare_new} Comparison between the theoretical response at higher strains with the experimental response of the samples sLN-y and 5Mg-LN-z. Asterisks label modes that are Raman silent in the given scattering geometry.}
	\begin{ruledtabular}
	
	\begin{tabular}{c|c c c c|c c c c}
	\hline
    &\multicolumn{4}{c|}{sLN-y (cm$^{-1}$/\%)}&\multicolumn{4}{c}{5Mg-LN-z (cm$^{-1}$/\%)}\\[1ex]

    Modes&$\mathrm{\overline{Z}}$(XX)Z & $\mathrm{\overline{Z}}$(XY)Z & $\mathrm{\overline{Z}}$(YY)Z & Theory &$\mathrm{\overline{X}}$(ZZ)X & $\mathrm{\overline{X}}$(YZ)X & $\mathrm{\overline{X}}$(YY)X & Theory\\
    \hline
	$\mathrm{A_1(LO_1)}$ & 8.01 $\pm$ 1.41& * & 6.31 $\pm$ 0.68 & * & * & * & * & *\\
    $\mathrm{A_1(LO_2)}$ & 7.95 $\pm$ 0.35 & * & 6.18 $\pm$ 0.68 & * & * & * & * & *\\ 
    $\mathrm{A_1(LO_3)}$ & 4.24 $\pm$ 1.94& * & 2.82 $\pm$ 0.68 & * & * &* & * & *\\
    $\mathrm{A_1(LO_4)}$ & 6.75 $\pm$ 0.33 & * & 5.12 $\pm$ 0.68 & * & * & * & * & * \\[1ex]
	$\mathrm{A_1(TO_1)}$ & * & * & * & 1.15 & -4.97 $\pm$ 0.93 & * & -5.10 $\pm$ 5.63 & -2.97\\
    $\mathrm{A_1(TO_2)}$ & * & * & * & 3.83 & 7.49 $\pm$ 1.08 & * & 5.98 $\pm$ 4.73 & 4.44\\
    $\mathrm{A_1(TO_3)}$ & * & * & * & 4.47 & 4.54 $\pm$ 1.24 & * & -0.19 $\pm$ 1.4 & 0.65\\
    $\mathrm{A_1(TO_4)}$ & * & * & * & 3.87 & -4.94 $\pm$ 0.65 & * & -6.94 $\pm$ 0.60 & -4.91\\[1ex]
    $\mathrm{E(TO_1)}$ & 6.27 $\pm$ 0.63 & 5.97 $\pm$ 0.63 & 4.41 $\pm$ 0.60 & 3.25/0.83 & * & 0.40 $\pm$ 0.44 & -0.15 $\pm$ 0.54 & -0.61\\
    $\mathrm{E(TO_2)}$ & 5.48 $\pm$ 0.35 & 4.84 $\pm$ 0.35 & 1.56 $\pm$ 0.34 & -0.70/2.36 & * & 0.61 $\pm$ 0.53 & 1.08 $\pm$ 1.51 & -2.55\\
    $\mathrm{E(TO_3)}$ & 7.56 $\pm$ 1.04 & 5.81 $\pm$ 1.04 & 4.68 $\pm$ 1.07 & 1.34/2.21 & * & 1.46 $\pm$ 1.75 & 1.44 $\pm$ 7.44 & 0.50\\
    $\mathrm{E(TO_4)}$ & 4.5 $\pm$ 0.69 & 4.44 $\pm$ 0.69 & 2.42$\pm$ 0.68 & 2.03/2.12 & * & 5.43 $\pm$ 0.92 & -3.84 $\pm$ 11.74 & 2.40\\
    $\mathrm{E(TO_{5})}$ & * & * & * & 6.73/6.55 & * & * & * & -0.68\\
    $\mathrm{E(TO_{6})}$ & 12.9 $\pm$ 1.85 & 11.13 $\pm$ 1.85& 9.16 $\pm$ 2.02 & 13.00/11.86 & * & 0.72 $\pm$ 2.31 & 1.90 $\pm$ 1.9 & -0.94\\
    $\mathrm{E(TO_7)}$ & 3.41 $\pm$ 0.68& 1.64 $\pm$ 0.68& 1.23 $\pm$ 0.82 & 4.12/2.01 & * & 7.21 $\pm$ 2.64 & 7.69 $\pm$ 0.59 & 4.75\\
    $\mathrm{E(TO_8)}$ & 8.56 $\pm$ 0.48 & 6.72 $\pm$ 0.48 & 3.27 $\pm$ 0.70 & 3.75/1.00 & * & 6.73 $\pm$ 0.62 & 6.20 $\pm$ 5.94 & 3.35 \\
    $\mathrm{E(TO_9)}$ & - & - & - & 5.41/0.91 & - & - & -7.95 $\pm$ 13.41 & 0.90\\
    \hline
	\end{tabular}

	\end{ruledtabular}
\end{table*}

\begin{figure}
	\includegraphics{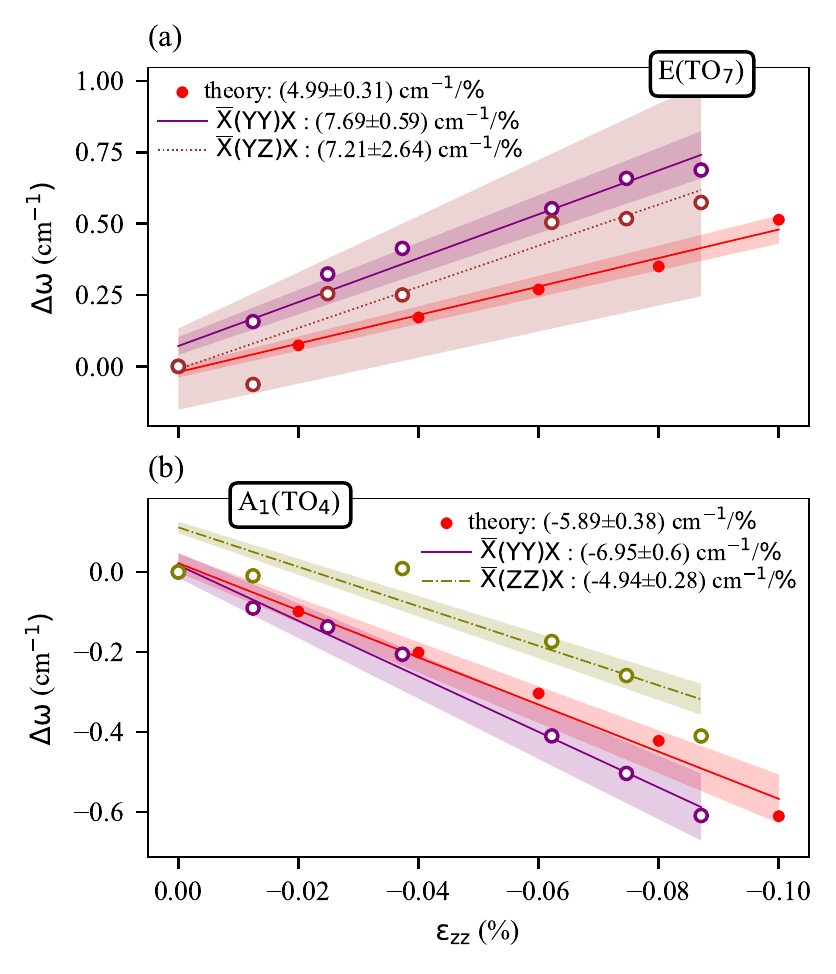}
	\caption{\label{fig:result_compare_z-compression}Comparison of theoretical stoichiometric and experimental 5Mg-LN-z sample when compressing along the z-direction, for frequency shifts of (a) E(TO$_7$) and (b) A$_1$(TO$_4$). E(TO$_7$) shows a positive slope while A$_1$(TO$_4$) shows a negative slope for z compression. } 
\end{figure}

As concluded from the experimental results in Sec.~\ref{sec:experimental_results}, the 5$\%$ MgO-doped LN responds equally well as the stoichiometric LN, upon compression. We compare the 5Mg-LN-z sample with theoretical data of stoichiometric LN for z compression. As theoretically predicted and confirmed by the experiment, both E(TO$_7$) and A$_1$(TO$_4$) modes show a linear dependence on z compression [see Fig.\ref{fig:result_compare_z-compression}(a)-(b)]. For the E(TO$_7$) mode, no degeneracy lifting occurs for z compression, which can also be seen in the experimental data. This investigation provides another rule for characterizing compressed LN.

As described in Sec.~\ref{sec:theoreticalResults} the frequencies of most modes are linearly increasing with increasing compression. For a few modes, this is not the case, e.g., for the A$_1$(TO$_4$) modes. For this mode, a linear decrease in frequency with increasing compression is theoretically predicted. This is consistent with experimental measurements, as shown in Fig.~\ref{fig:result_compare_z-compression}(b). This unusual behavior can also be explained with the help of the eigenvectors. In Fig.~\ref{fig:ev} it can be seen that the oxygen ions move towards the positively charged niobium ions and then away from them. If z compression reduces the distance between the cations, then as the oxygen ions displace away from niobium, the restoring force towards niobium may be weakened by the shortened distance to the lithium ions, which also exert the Coulomb force on the negatively-charged oxygen ions. Thus, the frequency for z compression might be reduced. In a similar way, this explanation can also be applied to A$_1$(TO$_1$) modes. All other modes for comparison for z compression are shown in Table~\ref{table:compare_new}. 

\section{Conclusions}
Experimental and theoretical phonon frequencies as a function of uniaxial strain have been evaluated and compared for LN and LT. This investigation has shown that all phonon modes are affected by strain. Both theoretical and experimental results are consistent with the roughly linear behavior of the modes and can be reasonably explained by an analysis of the calculated eigenvectors. The linearity of the slopes confirms the 
previous works under hydrostatic pressure and under external electric fields \cite{Stone2011,MendesFilho1984}. 

Accordingly, the obtained slopes provide reference values for the peak shifts. Furthermore, with the help of the slopes, it is possible to characterize compressed LN crystals and, hence, the vicinity of domain walls and wave guides in LN. Our experimental and theoretical investigations show that x and y compressed LN has particularly high peak shifts for the E(TO$_{5/6}$) modes and z compressed LN for the A$_1$(TO$_2$), E(TO$_7$) and E(TO$_8$) modes \cite{MendesFilho1984}.

Furthermore, calculations have shown that the degeneracy of the E modes is lifted for x and y compression upon symmetry reduction. The splitting has also been observed in our experiments when measured under different scattering geometries. Based on our theoretical and experimental investigations three rules can be formulated to characterize the direction of uniaxial compression of a LN sample using different scattering geometries. This only concerns the E modes. The sample is x compressed if higher frequency shifts are observed for the $\mathrm{\overline{Z}}$(YY)Z geometry than for the $\mathrm{\overline{Z}}$(XY)Z  and $\mathrm{\overline{Z}}$(XX)Z geometries when applying the same compressive stress. For y compression, the opposite is the case. In contrast, when no differences are observed between the frequency shift of the E modes for various scattering geometries, it can be concluded that the sample under investigation must be z compressed.

Furthermore, the calculation has shown that the behavior of LT under strain is not significantly different from that of LN. Experimentally, it has additionally revealed that the frequency shifts under compression of stoichiometric and 5\% MgO-doped LN are comparable. As DWs in LN and LT can be represented in the first approximation as strained lattices, our investigation might be helpful to understand the Raman signal of DWs. Further studies of strained samples, concerning e.g., their linear and non-linear optical response might further help to characterize ferroelectric DWs.

\begin{acknowledgments}
We thank Dr Clifford W. Hicks for the help in designing the stress cell used in the measurements. The authors gratefully acknowledge the financial support by the Deutsche Forschungsgemeinschaft (DFG) through joint DFG-ANR project TOPELEC (No. EN-434/41-1 and No. ANR-18-CE92-0052),
FOR5044 (ID: 426703838; \url{http:\\www.For5044.de}), as well as the Würzburg-Dresden Cluster of Excellence ”ct.qmat” (EXC 2147). Calculations for this research were conducted on the Lichtenberg high-performance computer of the TU Darmstadt and at the Höchstleistungrechenzentrum Stuttgart (HLRS). The authors furthermore acknowledge the computational re- sources provided by the HPC Core Facility and the HRZ of the Justus-Liebig-Universität Gießen. Furthermore, we thank D. Bieberstein and T. Gemming from IFW Dresden for assistance with dicing of the wafers and support from the International Max Planck Research School for Chemistry and Physics of Quantum Materials.

\end{acknowledgments}

\bibliographystyle{ieeetr}
\bibliography{raman_stress_paper_new}

\newpage
\foreach \x in {1,...,17}
{
	\clearpage
	\includepdf[pages={\x,{}}]{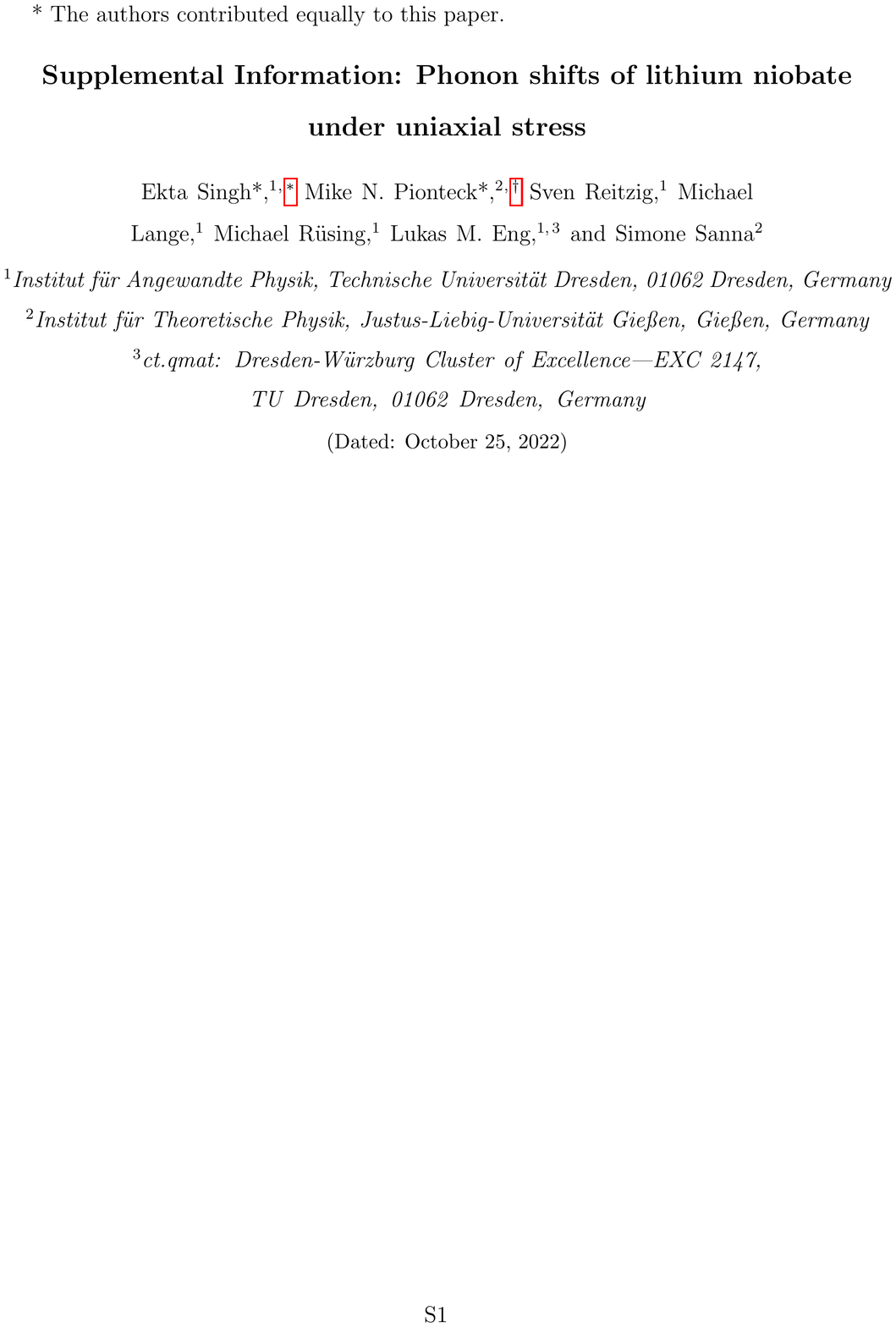}
}

\end{document}

\title{Supplemental Information: Phonon shifts of lithium niobate under uniaxial stress}

\author{Ekta Singh*}
\email[email]{: ekta.singh1@tu-dresden.de}
\affiliation{Institut für Angewandte Physik, Technische Universität Dresden, 01062 Dresden, Germany}

\author{Mike N. Pionteck*}
\email[email]{: mike.pionteck@theo.physik.uni-giessen.de}
\affiliation{Institut für Theoretische Physik, Justus-Liebig-Universität Gießen, Gießen, Germany}

\author{Sven Reitzig}
\affiliation{Institut für Angewandte Physik, Technische Universität Dresden, 01062 Dresden, Germany}

\author{Michael Lange}
\affiliation{Institut für Angewandte Physik, Technische Universität Dresden, 01062 Dresden, Germany}

\author{Michael Rüsing}
\affiliation{Institut für Angewandte Physik, Technische Universität Dresden, 01062 Dresden, Germany}

\author{Lukas M. Eng}
\affiliation{Institut für Angewandte Physik, Technische Universität Dresden, 01062 Dresden, Germany}
\affiliation{ct.qmat: Dresden-Würzburg Cluster of Excellence—EXC 2147, TU Dresden, 01062 Dresden, Germany}

\author{Simone Sanna}
\affiliation{Institut für Theoretische Physik, Justus-Liebig-Universität Gießen, Gießen, Germany}

* The authors contributed equally to this paper.
\date{\today}

\keywords{broadband coherent anti-Stokes Raman scattering, B-CARS, CARS, solid state, ferroelectrics, single crystals, lithium niobate, lithium tantalate, phase-retrieval, phase matching, selection rules, phonon assignment}
\maketitle

\section{Optimizing the measurements}\label{sec:optimisation}

In the stress cell, the samples are mounted onto two sample plates above a 1 mm gap. This gap is the region where the maximum strain is expected. The strain is gradually transmitted through the epoxy to the sample. Therefore, finding an optimum location on the samples is central for this work  to achieve reproducible results.

\begin{figure*}
	\includegraphics{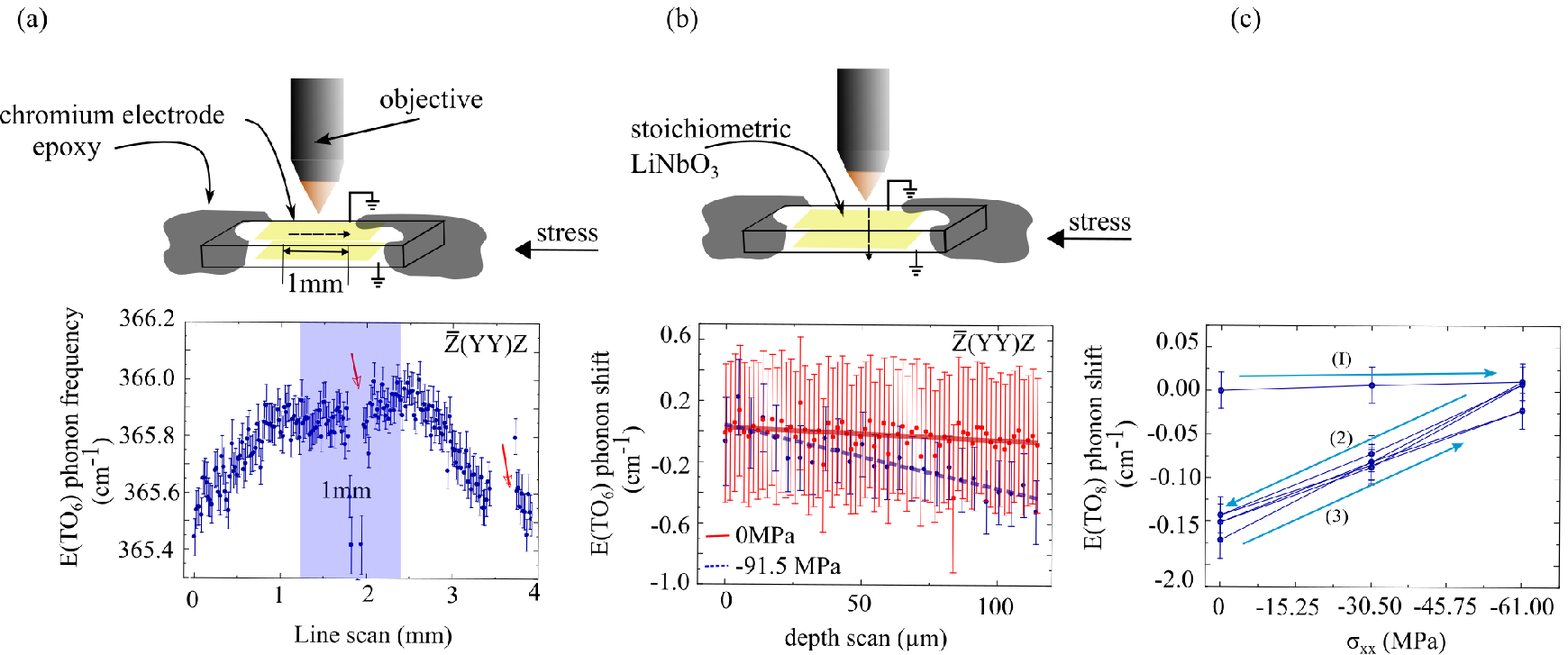}
	\caption{\label{fig:experimental_approach}Optimization of the measurement on sample sLN-x. (a) Line scan at -91.5 MPa. (b) depth scan on unstressed and compressed sLN-x based on the response of the E(TO$_6$) phonon frequency. (c) Hysteresis behavior of the E(TO$_8$) phonon mode with applied stress ($\sigma_{\mathrm{xx}}$) applied along the crystallographic x-axis during the first and following cycles. Here, after one initial cycle, a reproducible behavior was established. }
\end{figure*}

In order to do so, we performed a series of optimization measurements. With the help of Raman spectroscopy the distribution of the strain was measured along the length of the sample (i.e. the strained axes), see Fig.~\ref{fig:experimental_approach}(a). Here, a stoichiometric LN (sLN-x) sample was compressed to the stress of $\SI{-91.5}{\mathrm{MPa}}$ along the crystallographic x-axis. To see the gradual increase of the strain towards the center from both ends of the sample, the Raman spectra were measured at 200 points over a $\SI{4}{\mathrm{mm}}$ length of the sample with increments of $\SI{20}{\upmu\mathrm{m}}$. As the E(TO$_6$) phonon is one of the isolated peaks and also as demonstrated in the main text, this phonon shows the strongest response to the stress, we now show the peak frequency of the E(TO$_6$) as a function of position. Indeed, the plot shows that the frequency of E(TO$_6$) mode increases towards the center of the compressed sLN-x and stays constant over $\approx$1 mm length, indicating that the strain is transmitted to the sample through the epoxy and is uniform in the center of the sample. Based on this result we decided to measure always in the center of the $\SI{1}\mathrm{\text{mm}}$ gap of the device, which is easily identified in the optical microscope. One may also notice the two red colored arrows pointing to missing or shifted data points. This was observed in all the phonon modes. These likely indicate dust particles sitting on the sample surface. Such areas were avoided during subsequent measurements.

Furthermore, since the sample is sitting on top of the $\SI{1}{\mathrm{mm}}$ gap, in order to investigate a bending effect, depth scans on the same compressed sample at $\SI{-91.5}{\mathrm{MPa}}$ stress (sLN-x) were performed, and the results were compared to a reference unstressed stoichiometric LN sample, as shown in Fig.~\ref{fig:experimental_approach}(b). To achieve the depth scan, the laser was focused at different depths along the sample thickness. The first thing to notice is that the frequency of phonon mode E(TO$_6$) decreases with a slightly different slope when moving into the depth of both the unstressed and stressed samples. The error bars in the unstressed sample are of the same order of magnitude as the slope, indicating no strain inhomogeneity with depth for the unstressed sample. On the other hand for the sample sLN-x compressed at $\SI{-91.5}{\mathrm{MPa}}$, we observe a small but significant slope of $(-3.87\pm0.411)\times10^{-3}\mathrm{cm^{-1}\upmu m^{-1}}$, which can be interpreted in terms of the sample is bent downwards. Such slight bends may be caused by very small inhomogeneities in the mounting, variations in thickness, or width of the sample, and cannot be avoided completely. 
%there is a possibility of the sample bending. Depending on how the epoxy is placed on the edges of the sample or the slightest variations in sample thickness, it can bend upwards or downwards and as a result, can lead to an increase or decrease of the phonon frequency as a function of sample thickness. 

As each sample is mounted individually, it may bend slightly upwards or downwards, which will cause a slight offset error in the frequency, as additional uniaxial stress is added or subtracted. Assuming the determined depth dependency for this sample is an average slope for up- or downward bends, we can use this to estimate the error that is caused by this effect. Whenever a sample is bent, this will result in three distinct regions in the sample: 
\begin{itemize}
    \item At a line parallel to the horizontal center no bending strain will be experienced (neutral line).
    \item The region above will experience compressive (tensile) strain for a downward (upward) bending.
    \item In the region below the neutral line the behavior will be reversed.
    
\end{itemize}
Our measurements were always performed at a depth of approximately $\SI{10}{\mathrm{\upmu}}$m with respect to the top surface, as this region could be located easily. All the samples in this work had thicknesses in the range of $\SI{100}{\mathrm{\upmu}}$m. Therefore, our measured spot was conservatively estimated at about a $\SI{50}{\mathrm{\upmu m}}$ distance from the neutral line. Therefore, if the sample is bent up- or downwards, an additional frequency shift at each phonon is added. For example, for the E(TO$_6$) based on the depth dependency determined in the paragraph above this amounts to an offset of $0.19\pm\SI{0.021}{\mathrm{cm}^{-1}}$ at a stress of $\SI{91.5}{\mathrm{MPa}}$. If we now compare this value to the error in the peak frequency due to fitting, which for the E(TO$_6$) in the sample sLN-x, for example, is $\mathrm{\Delta\omega} = \SI{0.08}{\mathrm{cm}^{-1}}$ then this is approximate twice the error for the largest stresses. Please note, that this error will proportionally decrease for smaller applied forces, as the bending will proportionally decrease as well. However, compared to the stress response of the E(TO$_6$) [see Table II in the main paper] this error is still small. For other phonons, a similar relative error will be observed, as both the bending error and the stress-dependent shift are proportional.

%For the sake of consistency the measurements were taken at a constant depth of 10 $\upmu$m inside from the surface of each sample. Since the bending effect is stress dependent and laser focus is placed $\approx$ 50 $\upmu$m above the neutral line running through the center of the bent sample, this inhomogeneity at -91.5 MPa would be of 0.19 $\pm$ 0.021 cm$^{-1}$ (50 $\upmu$m)$^{-1}$. Conservatively assuming a general case that the samples can bend either up or down, the bending error can be added above and below to the existing error of ETO$_6$ mode at compression of -91.5 MPa which for $\mathrm{\overline{Z}(YY)Z}$ is 0.08 cm$^{-1}$. Keeping in mind that this value can not be applied to other phonon modes because they respond with different strength than E(TO$_6$), which also, as shown later is the mode which responses the most to the stress. For other phonon modes, the bending effect should be even less because other modes respond weaker to the stress compared to E(TO$_6$) [see table \ref{table:compare_new}]. As a result, the bending effect is small and can be neglected.

In addition to the strain inhomogeneity along the length and depth of the sample, we expect that the hardening of the epoxy (during curing) may put the sample under stress even before measurement, which may relax during measurement cycles. Indeed, we observed that some phonon modes such as E(TO$_8$) showed a hysteresis behavior. As seen from Fig.~\ref{fig:experimental_approach}(c), the first cycle shows hysteresis but typically after 3 cycles, the response was reproducible within error limits. As mentioned earlier this behavior is a result of epoxy and indirectly sample relaxation after the first cycle. Therefore, such a procedure was performed for each sample. 

In conclusion, these are three optimization steps that were always considered before starting any measurement on a new sample.

\newpage

\section{Young's modulus calculation for lithium niobate}\label{sec:Youngs modulus}
Mechanical properties of a solid are described by the fourth rank elastic stiffness tensor $c_{ijkl}$ and the elastic compliance tensor $s_{ijkl}$. $s_{ijkl}$ is the inverse of elastic stiffness tensor ($c_{ijkl}$), i.e. $s_{ijkl} = c_{ijkl}^{-1}$. Both these tensors connect the symmetric and second rank strain ($\epsilon_{ij}$) and stress ($\sigma_{ij}$) tensors to each other.  The stress is defined by force applied to a unit area of material, while the strain is the relative change of length in a material. The SI unit of the stress is given by Nm$^{-2}$, while a strain is a unitless quantity. In the elastic regime, $c_{ijkl}$ and $s_{ijkl}$ are connected to stress and the strain via Hooke's law:
\begin{equation}\label{eq:stress and strain equation 1}
\begin{split}
\epsilon_{ij} = s_{ijkl} \sigma_{kl}\\
\sigma_{ij} = c_{ijkl} \epsilon_{kl}	
\end{split}
\end{equation}

Equation (\ref{eq:stress and strain equation 1}) is written in a tensorial form, which can be simplified into matrix form by using Voigt notation, where indices $i$, $j$, $k$, and $l$ can take values from 1, 2, and 3 which correspond to x, y, and z directions, respectively.

11 $\rightarrow$ 1 $\hfil$ 12 $\rightarrow$ 6 $\hfil$ 13 $\rightarrow$ 5 \\
22 $\rightarrow$ 2 $\hfil$ 21 $\rightarrow$ 6 $\hfil$ 23 $\rightarrow$ 4 \\
33 $\rightarrow$ 3 $\hfil$ 31 $\rightarrow$ 5 $\hfil$ 32 $\rightarrow$ 4 \\

Since, the strain and the stress tensors are symmetric (ij = ji), the 81 components  (3$^4$) of the fourth rank stiffness and compliance tensor are reduced to 36 independent components, where $c_{ijkl} = c_{jikl} = c_{jilk} = c_{{ijlk}}$. It is due to this fact that the elastic constants can be written in 6 $\times$ 6 matrix. This matrix can be further simplified to 21 independent components because the stiffness coefficients are second-order derivatives of the mechanical energy density with respect to strain components. The order in which the differentiation is taken does not affect the energy density, which makes the stiffness coefficients symmetric as well.

Furthermore, the symmetries of a crystal can simplify these matrices even more with the help of Neumann's law~\cite{Newnham2004}. In the case of ferroelectric LiNbO$_3$ crystal, which belongs to the trigonal crystal system and 3m point group, the final form of the matrices is given as follows \cite{Newnham2004, Weis1985}. 

\textbf{Elastic stiffness matrix of LiNbO$_3$} 

\[
\begin{bmatrix}{\label{matrix:stiffness}}

c_{11}&c_{12}&c_{13}&c_{14}&0&0\\
c_{12}&c_{22}&c_{13}&-c_{14}&0&0\\
c_{13}&c_{13}&c_{33}&0&0&0\\
c_{14}&-c_{14}&0&c_{44}&0&0\\
0&0&0&0&c_{44}&c_{14}\\
0&0&0&0&c_{14}&\dfrac{1}{2}(c_{11}-c_{12})\\	

\end{bmatrix}
\]

\textbf{Elastic compliance matrix of LiNbO$_3$}
\[
\begin{bmatrix}{\label{matrix:compliance}}
s_{11}&s_{12}&s_{13}&s_{14}&0&0\\
s_{12}&s_{22}&s_{13}&-s_{14}&0&0\\
s_{13}&s_{13}&s_{33}&0&0&0\\
s_{14}&-s_{14}&0&s_{44}&0&0\\
0&0&0&0&s_{44}&2s_{14}\\
0&0&0&0&2s_{14}&2(s_{11}-s_{12})\\	

\end{bmatrix}
\]

\begin{table}[h]
\caption{\label{table:coeff_table}The room temperature value of elastic compliance and stiffness coefficients at the constant electric field for LiNbO$_3$. }
\centering
\begin{tabular}{|c|c|c|c|c|c|c|}
\hline
Compliance coefficients ($\times 10^{-11}$\ m$^{2}$N$^{-1}$ )& s$_{11}$ & s$_{12}$ & s$_{13}$ & s$_{14}$ & s$_{33}$ & s$_{44}$\\
	\cite{Warner1967a} & 0.578 & -0.101 & -0.147 & -0.102 & 0.502 & 0.170\\
	\hline
Stiffness coefficients ($\times 10^{11}$\ Nm$^{-2}$ )& c$_{11}$ & c$_{12}$ & c$_{13}$ & c$_{14}$ & c$_{33}$ & c$_{44}$\\	
	\cite{Warner1967a}& 2.03& 0.53& 0.75& 0.09& 2.45& 0.60\\
	\hline

\end{tabular} 

\end{table}

The values of these coefficients are measured by Warner et.al., provided in table~\ref{table:coeff_table}. However, the stoichiometry of LiNbO$_3$ is not provided by them in that paper~\cite{Warner1967a}. Assuming that the values of $c_{ijkl}$ coefficients do not change dramatically for the 5$\%$ MgO-doped and stoichiometric LiNbO$_3$, the values given in table \ref{table:coeff_table} have been used in this thesis for further calculations of the Young's Modulus. Using the matrix form of equation (\ref{eq:stress and strain equation 1}), the stress and the strain can be written as equation (\ref{eq:stress and strain equation 2}) and (\ref{eq:stress and strain equation 3}) as follows:
\begin{equation}\label{eq:stress and strain equation 2}
\begin{split}
    \epsilon_1 = s_{11}\sigma_1 + s_{12}\sigma_2 + s_{13}\sigma_3 + s_{14}\sigma_4\\
	\epsilon_2 = s_{12}\sigma_1 + s_{22}\sigma_2 + s_{13}\sigma_3-s_{14}\sigma_4\\
	\epsilon_3 = s_{13}\sigma_1 + s_{13}\sigma_2 + s_{33}\sigma_3\\
	\epsilon_4 = s_{14}\sigma_1 - s_{14}\sigma_2+s_{44}\sigma_4
\end{split}
\end{equation}

and

\begin{equation}\label{eq:stress and strain equation 3}
\begin{split}
    \sigma_1 = c_{11}\epsilon_1 + c_{12}\epsilon_2 + c_{13}\epsilon_3 + c_{14}\epsilon_4\\
	\sigma_2 = c_{12}\epsilon_1 + c_{22}\epsilon_2 + c_{13}\epsilon_3-c_{14}\epsilon_4\\
	\sigma_3 = c_{13}\epsilon_1 + c_{13}\epsilon_2 + c_{33}\epsilon_3 \\
	\sigma_4 = c_{14}\epsilon_1 - c_{14}\epsilon_2 + c_{44}\epsilon_4
\end{split}
\end{equation}

\begin{figure*}[!h]
\centering
\includegraphics{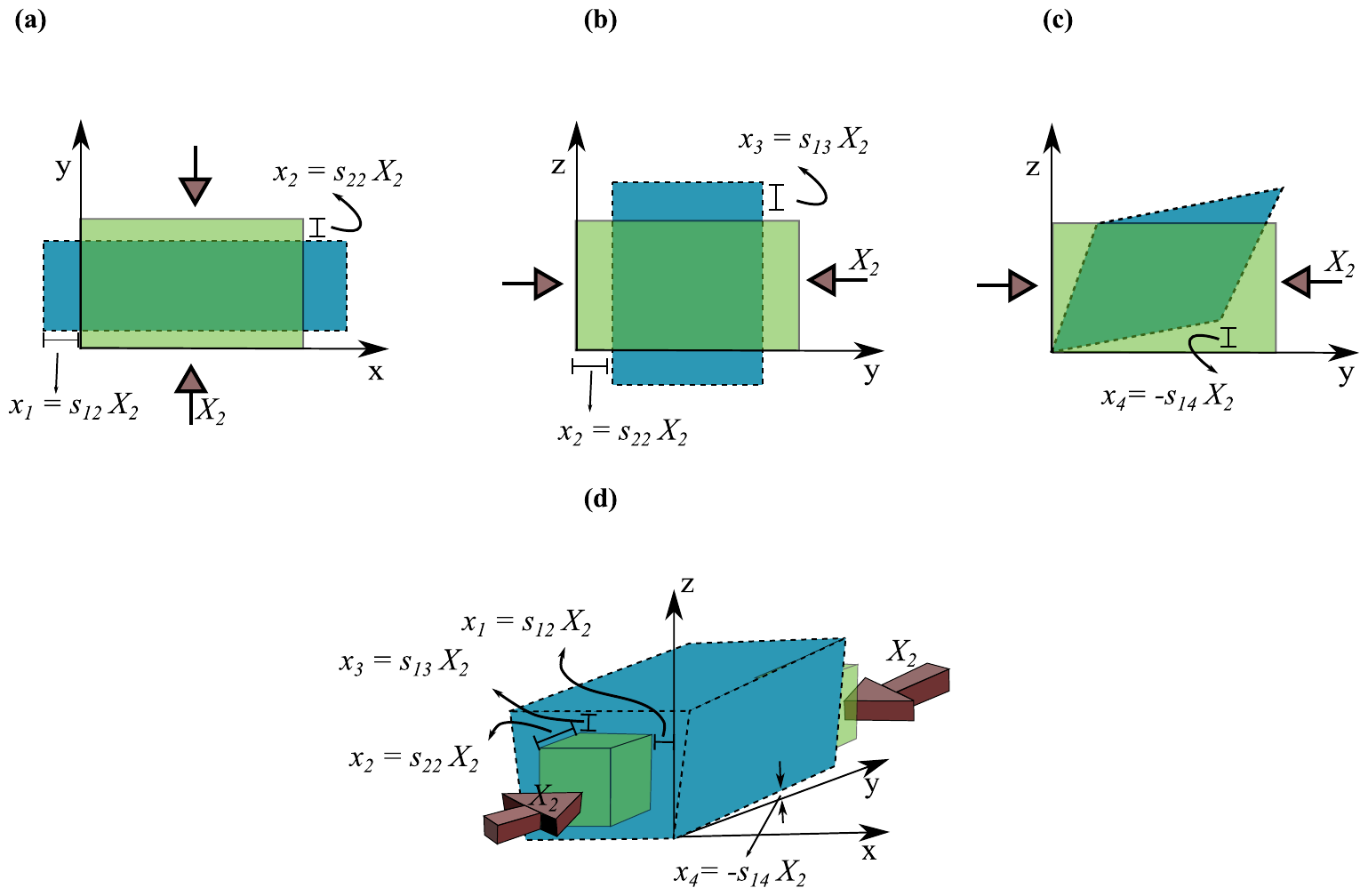}
\caption{\label{fig:stress_and_strain}Distribution of the strain according to equation~\ref{eq:stress and strain equation 2}, when the compressive stress is applied only along y-axis in LiNbO$_3$ ($\sigma_1=0, \sigma_2 \neq 0, \sigma_3=0$). Green color represents unstrained and blue color represents the strained shape of the rectangular shape crystal according to the equations (\ref{eq:stress and strain equation 2}) : (a) Normal strain $\epsilon_1$ and $\epsilon_2$ alone. (b) Normal strain $\epsilon_3$ alone. (c) Shear strain $\epsilon_4$ alone. (d) All the strains from (a), (b), and (c) are combined together in a 3-dimensional image.}
\end{figure*}

In order to get the Young's Modulus along x-axis, we need the ratio $E_{11} = \dfrac{\sigma_1}{\epsilon_1}$, which can be obtained by dividing $\sigma_1$ in equation (\ref{eq:stress and strain equation 3}) by $\epsilon_1$, as follows.
\begin{equation}\label{eq:youngs modulus x}
	E_{11} = \dfrac{\sigma_1}{\epsilon_1} = c_{11} + c_{12}\dfrac{\epsilon_2}{\epsilon_1} + c_{13}\dfrac{\epsilon_3}{\epsilon_1} + c_{14}\dfrac{\epsilon_4}{\epsilon_1} 
\end{equation}
 
 where $\dfrac{\epsilon_2}{\epsilon_1}$, $\dfrac{\epsilon_3}{\epsilon_1}$ or $\dfrac{\epsilon_4}{\epsilon_1}$ are quantities equal to the negative of Poisson's ratio. These ratios can also be obtained by using equation (\ref{eq:stress and strain equation 2}). For uniaxial stress along the x-axis, all the stress components except $\sigma_{1}$ should be zero. Using this information one can calculate the Poisson's ratio of LiNbO$_3$ as follows: 
 \begin{equation}
 	\begin{split}
 		\nu_{12} = -\dfrac{\epsilon_2}{\epsilon_1} = -\left( \dfrac{s_{12}}{s_{11}} \right) = -\left( \dfrac{-1.01}{5.78} \right) = 0.1747\\
 		\nu_{13} = -\dfrac{\epsilon_3}{\epsilon_1} = -\left( \dfrac{s_{13}}{s_{11}} \right) = -\left( \dfrac{-1.47}{5.78} \right) = 0.254\\
 		\nu_{14} = -\dfrac{\epsilon_4}{\epsilon_1} = -\left( \dfrac{s_{14}}{s_{11}} \right) = -\left( \dfrac{-1.02}{5.78} \right) = 0.17647
 	\end{split}
 \end{equation}
  
Inserting these values in equation (\ref{eq:youngs modulus x}) one can calculate Young's Modulus along the x-axis. The same method of calculation is applied to calculate the Young's Modulus of LiNbO$_3$ along the y and z axes, for values see table \ref{table:Young's modulus table}. 

\begin{table}[h]
\centering
\caption{\label{table:Young's modulus table}Calculated Young's Modulus of LiNbO$_3$ along the x, y, and z crystallographic axes}
\begin{tabular}{|c|c|c|}
\hline
    $E_{11}$ (GPa) & $E_{22}$ (GPa)& $E_{33}$ (GPa)\\
	\hline
	173.6 & 173.07 & 201.0\\
	\hline
\end{tabular}
\end{table}

\clearpage

\section{Results}
\subsection{Experiment}
In this section all the experimental slopes of compressed samples sln-x/y, 5Mg-LN-x/y/z are provided in the tables~\ref{tab:sln_exp_xy},~\ref{tab:5Mg_exp_xy}, and ~\ref{tab:5Mg_exp_z} respectively. Table~\ref{tab:sln_exp_tens_x} provides experimental slopes of the tensile stress on sample sLN-x only.
\begin{table}[h]
    \caption{\label{tab:sln_exp_xy} Experimental phonon frequency shift values $\mathrm{\Delta\omega}$ of stoichiometric lithium niobate upon uniaxial compression along the x- and y-axis.}
    
\begin{ruledtabular}
\begin{tabular}{ c|c c c|c c c}
\hline
&\multicolumn{3}{c|}{sLN-x (cm$^{-1}$/GPa)}&\multicolumn{3}{c}{sln-y (cm$^{-1}$/GPa)}\\[1ex]
Phonon modes&$\mathrm{\overline{Z}}$(XX)Z & $\mathrm{\overline{Z}}$(XY)Z & $\mathrm{\overline{Z}}$(YY)Z &$\mathrm{\overline{Z}}$(XX)Z & $\mathrm{\overline{Z}}$(XY)Z & $\mathrm{\overline{Z}}$(YY)Z\\
\hline
$\mathrm{A_1(LO_1)}$ & 2.13 $\pm$ 0.52 & * & 2.06 $\pm$ 0.58 & 4.61 $\pm$ 0.81 & * & 3.64 $\pm$ 1.89\\
$\mathrm{A_1(LO_2)}$ & 3.29 $\pm$ 0.17 & * & 3.66 $\pm$ 0.17 & 4.58 $\pm$ 0.20 & * & 3.56 $\pm$ 0.60\\
$\mathrm{A_1(LO_3)}$ & 0.39 $\pm$0.94 & * & 0.29 $\pm$ 0.71  & 2.44 $\pm$  1.11&* & 1.62 $\pm$ \\
$\mathrm{A_1(LO_4)}$ & 1.96 $\pm$ 0.21 & * & 2.32 $\pm$ 0.19 & 3.89 $\pm$ 0.19 & * & 2.95 $\pm$ 0.39 \\
\hline
$\mathrm{E(TO_1)}$ & 0.7 $\pm$ 0.26 & 0.63 $\pm$ 0.26 & 1.98 $\pm$ 0.27 & 3.62 $\pm$ 0.36 &  3.44 $\pm$ 0.36 & 2.54 $\pm$ 0.35 \\
$\mathrm{E(TO_2)}$ & 0.34 $\pm$ 0.17 & 0.26 $\pm$ 0.25 & 1.6 $8\pm$ 0.16  & 3.16 $\pm$ 0.20 & 2.79 $\pm$ 0.20 & 0.9 $\pm$ 0.20 \\
$\mathrm{E(TO_3)}$ & 1.28 $\pm$ 0.44 & 1.39 $\pm$ 0.65 & 1.67 $\pm$ 0.42 & 4.35 $\pm$ 0.6& 3.35 $\pm$ 0.6 & 2.7 $\pm$ 0.62\\
$\mathrm{E(TO_4)}$ & 2.12 $\pm$ 0.35 & 1.56 $\pm$ 0.59 & 3.02 $\pm$ 0.4 & 
 2.59 $\pm$ 0.4 & 2.56 $\pm$ 0.4 & 1.39 $\pm$ 0.39\\
$\mathrm{E(TO_{5/6})}$ & 6.36 $\pm$ 0.98 & 6.25 $\pm$ 1.5 & 8.17 $\pm$ 0.87 & 7.43 $\pm$ 1.06 & 6.41 $\pm$ 1.06 & 5.28 $\pm$ 1.16\\
$\mathrm{E(TO_7)}$ & 0.44 $\pm$ 0.37 & 0.86 $\pm$ 1.51 & 1.24 $\pm$ 0.22 & 1.97 $\pm$ 0.39 & 0.94 $\pm$ 0.39 & 0.71 $\pm$ 0.47\\
$\mathrm{E(TO_8)}$ & -0.47 $\pm$ 0.3  & 0.37 $\pm$ 0.36 & 3.03 $\pm$  0.27 & 4.93 $\pm$ 0.27 & 3.87 $\pm$ 0.27 & 1.89 $\pm$ 0.41\\
\hline
\end{tabular}
\end{ruledtabular}
\end{table}

\begin{table}[h]
    \caption{\label{tab:sln_exp_tens_x}Experimental phonon frequency shift values $\mathrm{\Delta\omega}$ of stoichiometric lithium niobate upon uniaxial tension along the x-axis.}
\begin{ruledtabular}
\begin{tabular}{|c|c|c|c|c|c|c|}
\multicolumn{7}{|c|}{Tension sLN-x (cm$^{-1}$/GPa),   Z(YY)Z}\\
\hline
A$_1$(LO$_1$)& A$_1$(LO$_2$) &A$_1$(LO$_3$)&A$_1$(LO$_4$)&E(TO$_1$)&E(TO$_2$)&E(TO$_3$)\\
-3.92 $\pm$ 2.39 &-2.35 $\pm$ 0.57 & 0.35 $\pm$ 2.6 & -1.2$\pm$0.55 &-4.04 $\pm$ 0.93 &-2.01$\pm$ 0.6&-3.01$\pm$1.72\\
\hline
E(TO$_4$)&E(TO$_{5/6}$)&E(TO$_7$)&E(TO$_8$)&&&\\
-2.7 $\pm$1.35 &-8.15 $\pm$ 2.88 &-0.86 $\pm$ 0.81 & -6.01 $\pm$0.85&&&
 \end{tabular}
 \end{ruledtabular}
 \end{table}

 \begin{figure*}[t]
	\centering
    \includegraphics{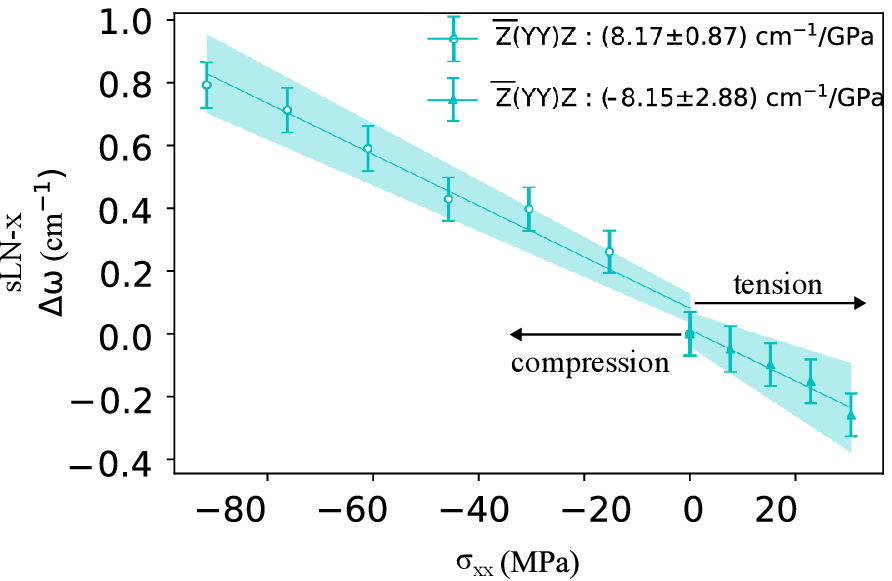}
    \caption{\label{fig:tens_comp} E(TO$_6$) phonon response under compression and tension along the x-axis in $\mathrm{\overline{Z}}$(YY)Z scattering geometry. }
	
\end{figure*}

\begin{table}[h]
    \caption{\label{tab:5Mg_exp_xy} Experimental phonon frequency shift values $\mathrm{\Delta\omega}$ of 5$\%$ MgO-doped lithium niobate upon uniaxial stress along the x- and y-axis.In the case of 5$\%$ MgO-doped LN, many peaks overlap with each other, therefore extracting the peak frequency with an acceptable error was not possible. This is represented by '-' sign.}
    
\begin{ruledtabular}
\begin{tabular}{ c|c c c|c c c}
\hline
&\multicolumn{3}{c|}{5Mg-LN-x (cm$^{-1}$/GPa)}&\multicolumn{3}{c}{5Mg-LN-y (cm$^{-1}$/GPa)}\\[1ex]
Phonon modes&$\mathrm{\overline{Z}}$(XX)Z & $\mathrm{\overline{Z}}$(XY)Z & $\mathrm{\overline{Z}}$(YY)Z &$\mathrm{\overline{Z}}$(XX)Z & $\mathrm{\overline{Z}}$(XY)Z & $\mathrm{\overline{Z}}$(YY)Z\\
\hline

$\mathrm{A_1(LO_1)}$ & - & * & - & -& * & -\\
$\mathrm{A_1(LO_2)}$ & - & * & - & -& * & -\\
$\mathrm{A_1(LO_3)}$ & - & * & - & -& * & -\\
$\mathrm{A_1(LO_4)}$ & 3.07 $\pm$ 0.83 & * & 3.79 $\pm$ 0.8 & 2.49 $\pm$ 0.42& * & 2.09 $\pm$ 0.45\\
\hline
$\mathrm{E(TO_1)}$ & 0.71 $\pm$ 0.35 & 1.05 $\pm$ 0.22 & 2.68 $\pm$ 0.29 & 1.88 $\pm$ 0.14& 1.85 $\pm$ 0.22 & 1.56 $\pm$ 0.18\\
$\mathrm{E(TO_2)}$ & -0.51 $\pm$ 0.45 & -0.63 $\pm$ 0.52 & 1.52 $\pm$ 0.3 & 1.64 $\pm$ 0.22 & 0.77 $\pm$ 0.33 & 0.83 $\pm$ 0.3 \\
$\mathrm{E(TO_3)}$ & - & - & - & -& - & -\\
$\mathrm{E(TO_4)}$ & - & - & - & -& - & -\\
$\mathrm{E(TO_{5/6})}$ & 0.71 $\pm$ 0.35 & 1.05 $\pm$ 0.2 2& 2.68 $\pm$ 0.29 & 1.88 $\pm$ 0.14 & 1.85 $\pm$  0.22 & 1.56 $\pm$ 0.18\\
$\mathrm{E(TO_7)}$ & - & - & - & -& - & -\\
$\mathrm{E(TO_8)}$ & -0.29 $\pm$ 0.8 & 1.66 $\pm$ 0.69 & 4.32 $\pm$ 0.67 & 2.46 $\pm$ 0.39 & 2.23 $\pm$ 0.59 & 1.06 $\pm$ 0.48\\

\hline

 \end{tabular}
 \end{ruledtabular}
 \end{table}

\begin{table}[h]
    \caption{\label{tab:5Mg_exp_z} Experimental phonon frequency shift values $\mathrm{\Delta\omega}$ of 5$\%$ MgO-doped lithium niobate upon uniaxial compression along the z-axis.}
\begin{ruledtabular}
\begin{tabular}{ c|c c c}
\hline
&\multicolumn{2}{c}{5Mg-LN-z (cm$^{-1}$/GPa)}\\[1ex]
Phonon modes&$\mathrm{\overline{X}}$(ZZ)X & $\mathrm{\overline{X}}$(YZ)X & $\mathrm{\overline{X}}$(YY)X\\
\hline

$\mathrm{A_1(TO_1)}$ & -2.47 $\pm$ 0.63 & * & -2.53 $\pm$ 2.8 \\
$\mathrm{A_1(TO_2)}$ & 3.72 $\pm$ 0.70  & * & 2.97 $\pm$ 2.17 \\
$\mathrm{A_1(TO_3)}$ & 2.23 $\pm$ 4.44  & * & -0.09 $\pm$ 0.70\\
$\mathrm{A_1(TO_4)}$ & -2.46 $\pm$ 0.14  & * & -3.46 $\pm$ 0.30 \\
\hline
$\mathrm{E(TO_1)}$ & * & 0.2 $\pm$ 0.22 & -0.07 $\pm$ 0.27 \\
$\mathrm{E(TO_2)}$ & * & 0.3 $\pm$ 0.26 & 0.54 $\pm$ 0.75 \\
$\mathrm{E(TO_3)}$ & * & 0.72 $\pm$ 0.87 & 0.07 $\pm$ 0.37 \\
$\mathrm{E(TO_4)}$ & * & 2.7 $\pm$ 0.46 & -1.91 $\pm$ 5.84 \\
$\mathrm{E(TO_{5/6})}$ & * & 0.36 $\pm$ 1.15 & 0.94 $\pm$ 3.19 \\
$\mathrm{E(TO_7)}$ & * & 3.58 $\pm$ 1.31  & 3.83 $\pm$ 0.3\\
$\mathrm{E(TO_8)}$ & * & 3.38 $\pm$ 0.31 & 3.08 $\pm$ 2.95 \\
\hline

 \end{tabular}
 \end{ruledtabular}
 \end{table}

\clearpage

\subsection{Theory}
In order to compare the calculated phonon frequencies in unstrained LN and LT with Ref.  \cite{Sanna2015}, they are displayed in Table \ref{tab:tabelComp1} and \ref{tab:tabelComp2}. In addition, Table \ref{table:ln_theory} and  \ref{table:lt_theory} show all the slopes of LN and LT under uniaxial compressive and tensile strain.
\begin{table}[h]
    \caption{Calculated frequencies of the Raman active phonon modes in unstrained LN in comparison with calculated and measured frequencies of Ref. \cite{Sanna2015}.}
        \label{tab:tabelComp1}
    \begin{ruledtabular}
    \begin{tabular}{ c c c c }
  Phonon mode & Theory $\mathrm{[cm^{-1}]}$ & Theory  $\mathrm{[cm^{-1}]}$ \cite{Sanna2015} & Exp. $\mathrm{[cm^{-1}]}$ \cite{Sanna2015}\\
\hline

$\mathrm{A_1(TO_1)}$ & 242 & 239 & 252–255 \\
$\mathrm{A_1(TO_2)}$ & 282 & 289 & 275–276 \\
$\mathrm{A_1(TO_3)}$ & 350 & 353 & 333–334 \\
$\mathrm{A_1(TO_4)}$ & 613 & 610 & 633 \\[1ex]
$\mathrm{E(TO_1)}$ & 150 & 148 & 150–151 \\
$\mathrm{E(TO_2)}$ & 216 & 216 & 237 \\
$\mathrm{E(TO_3)}$ & 265 & 262 & 262–263 \\
$\mathrm{E(TO_4)}$ & 319 & 323 & 320–321 \\
$\mathrm{E(TO_5)}$ & 372 & 380 & 367–369 \\
$\mathrm{E(TO_6)}$ & 384 & 391 & 367–369 \\
$\mathrm{E(TO_7)}$ & 420 & 423 & 432 \\
$\mathrm{E(TO_8)}$ & 577 & 579 & 580–581 \\
$\mathrm{E(TO_9)}$ & 668 & 667 & 664 \\

 \end{tabular}
 \end{ruledtabular}
 \end{table}
 
\begin{table}[h]
    \caption{Calculated frequencies of the Raman active phonon modes in unstrained LT in comparison with calculated and measured frequencies of Ref. \cite{Sanna2015}.}
        \label{tab:tabelComp2}
 \begin{ruledtabular}
 \begin{tabular}{ c c c c }
  Phonon mode & Theory $\mathrm{[cm^{-1}]}$ & Theory  $\mathrm{[cm^{-1}]}$ \cite{Sanna2015} & Exp. $\mathrm{[cm^{-1}]}$ \cite{Sanna2015}\\
\hline

$\mathrm{A_1(TO_1)}$ & 200 & 209 & 209–210 \\
$\mathrm{A_1(TO_2)}$ & 255 & 286 & 256–257 \\
$\mathrm{A_1(TO_3)}$ & 369 & 376 & 359–360 \\
$\mathrm{A_1(TO_4)}$ & 578 & 591 & 600 \\[1ex]
$\mathrm{E(TO_1)}$ & 139 & 144 & 143 \\
$\mathrm{E(TO_2)}$ & 193 & 199 & 210 \\
$\mathrm{E(TO_3)}$ & 247 & 253 & 254–257 \\
$\mathrm{E(TO_4)}$ & 313 & 319 & 315–317 \\
$\mathrm{E(TO_5)}$ & 370 & 409 & 383-384 \\
$\mathrm{E(TO_6)}$ & 384 & 420 & 383-384 \\
$\mathrm{E(TO_7)}$ & 452 & 459 & 460-465 \\
$\mathrm{E(TO_8)}$ & 579 & 590 & 592 \\
$\mathrm{E(TO_9)}$ & 658 & 669 & 661-662 \\

 \end{tabular}
 \end{ruledtabular}
 \end{table}

\begin{table}[h]
    \centering
    \caption{\label{table:ln_theory}Calculated slopes of transversal A$_1$ and E modes of LN under compressive and tensile strain in the x, y and z direction for strains in steps of $\SI{0.2}{\%}$ in $\mathrm{{cm}^{-1}}/\%$. The slopes for strain are defined as $\mathrm{(\Delta\omega)}/(\mathrm{|\Delta\epsilon|)}$.}
    \begin{ruledtabular}
	
    \begin{tabular}{c|c c c|c c c}
    \hline
    &\multicolumn{3}{c|}{compressive strain (cm$^{-1}$/\%)}&\multicolumn{3}{c}{tensile strain (cm$^{-1}$/\%)}\\[1ex]

    Phonon modes&x direction & y direction & z direction &x direction & y direction & z direction\\
	\hline
	$\mathrm{A_1(TO_1)}$ & 1.31 & 1.15 & -2.97 & -2.87 & -2.76 & 2.59\\
    $\mathrm{A_1(TO_2)}$ & 3.64 & 3.83 & 4.44 & -4.78 & -5.04 & -5.20\\
    $\mathrm{A_1(TO_3)}$ & 5.18 & 4.47 & 0.65 & -7.08 & -5.76 & -0.32\\
    $\mathrm{A_1(TO_4)}$ & 3.95 & 3.87 & -4.91 & -3.79 & -3.61 & 5.36\\[1ex]
    \hline
    $\mathrm{E(TO_1)}$ & 0.68/3.26 & 3.25/0.83 & -0.61 & -1.56/-3.37 & -3.58/-1.57 & 0.30\\
    $\mathrm{E(TO_2)}$ & 1.52/0.32 & -0.70/2.36 & -2.55 & -3.69/-0.04 & -1.19/-1.51 & 1.14\\
    $\mathrm{E(TO_3)}$ & 1.09/2.24 & 1.34/2.21 & 0.50 & -4.40/-1.56 & -2.99/-3.06 & -0.53\\
    $\mathrm{E(TO_4)}$ & 2.19/1.98 & 2.03/2.12 & 2.40 & -3.42/-1.65 & -1.55/-3.29 & -2.56\\
    $\mathrm{E(TO_{5})}$ & 6.58/6.09 & 6.73/6.55 & -0.68 & -9.57/-13.57& -13.97/-11.17 & 1.29\\
    $\mathrm{E(TO_{6})}$ & 12.84/12.39 & 13.00/11.86 & -0.94& -4.93/-3.76 & -2.57/-5.68 & 1.12\\
    $\mathrm{E(TO_7)}$ & 4.18/2.72 & 4.12/2.01 & 4.75 & 1.01/-1.52 & 0.98/-1.43 & -4.08\\
    $\mathrm{E(TO_8)}$ & 3.63/0.91 & 3.75/1.00 & 3.35 & -0.67/-4.51 & -1.17/-4.33 & -2.57\\
    $\mathrm{E(TO_9)}$ & 1.36/4.94 & 5.41/0.91 & 0.90 & 0.70/-5.12 & -4.67/0.19 & -0.52\\
    \hline
	\end{tabular}

	\end{ruledtabular}
\end{table}

\begin{table}[h]
    \centering
    \caption{\label{table:lt_theory}Calculated slopes of transversal A$_1$ and E modes of LT under compressive and tensile strain in the x, y and z direction for strains in steps of $\SI{0.4}{\%}$ in $\mathrm{cm}^{-1}$/\%. The slopes for strain are defined as $\mathrm{(\Delta\omega)}/(\mathrm{|\Delta\epsilon|)}$.}
	\begin{ruledtabular}
	
	\begin{tabular}{c|c c c|c c c}
	\hline
    &\multicolumn{3}{c|}{compressive strain (cm$^{-1}$/\%)}&\multicolumn{3}{c}{tensile strain (cm$^{-1}$/\%)}\\[1ex]

    Phonon modes&x direction & y direction & z direction &x direction & y direction & z direction\\
	\hline
	$\mathrm{A_1(TO_1)}$ & 3.16 & 3.11 & -2.78 & -2.40 & -2.41 & 2.58\\
    $\mathrm{A_1(TO_2)}$ & 4.51 & 4.43 & 3.77 & -3.66 & -3.50 & -3.93\\
    $\mathrm{A_1(TO_3)}$ & 4.84 & 3.87 & 1.71 & 1.51 & -6.90 & -1.60\\
    $\mathrm{A_1(TO_4)}$ & 3.98 & 1.92 & -4.93 & -4.64 & -4.28 & 4.93\\[1ex]
    \hline
    $\mathrm{E(TO_1)}$ & 1.32/1.45 & 1.82/0.98 & -0.49 & -0.55/-2.28 & -1.87/-1.03 & 0.45\\
    $\mathrm{E(TO_2)}$ & 2.68/0.91 & 0.22/3.45 & -3.36 & -3.96/-0.59 & -1.84/-2.91 & 2.00\\
    $\mathrm{E(TO_3)}$ & 2.62/2.45 & 2.79/2.40 & -1.35 & -2.19/-4.51 & -4.18/-2.85 & 1.21\\
    $\mathrm{E(TO_4)}$ & 2.73/2.40 & 2.88/2.18 & 1.59 & -3.75/-3.07 & -2.85/-3.62 & -1.97\\
    $\mathrm{E(TO_{5})}$ & 7.39/8.31 & 9.28/7.34 & -0.42 & -5.70/-15.88 & -16.06/-11.51 & 0.63\\
    $\mathrm{E(TO_{6})}$ & 14.82/13.58 & 14.46/14.06 & -0.97& -8.04/-5.76 & -4.54/-9.32 & 1.14\\
    $\mathrm{E(TO_7)}$ & 0.83/2.02 & 2.12/0.72 & 4.32 & -0.12/-1.35 & -1.21/-0.21 & -4.60\\
    $\mathrm{E(TO_8)}$ & 4.65/1.54 & 4.99/3.26 & 5.28 & -2.06/-1.25 & -0.80/-3.01 & -4.50\\
    $\mathrm{E(TO_9)}$ & 4.06/3.45 & 4.05/3.41 & 3.18 & -2.37/-3.26 & -2.50/-3.09 & -2.77\\
    \hline
    \end{tabular}

    \end{ruledtabular}
\end{table}
\clearpage

\bibliography{supp_raman-stress.bib}